\documentclass[useAMS,usegraphicx]{mn2e}
\usepackage{graphicx}
\usepackage{rotating}
\usepackage{mathptm}
\def\farcs{ \hbox{$.\!\!^{\prime\prime}$}}
\def\farcsec{\hbox{$.\!\!^{\rm s}$}}

\title[NIR/Optical Observations of Hotspots in Radio Galaxies]{NIR/Optical Counterparts of Hotspots in Radio Galaxies}

\author[K.-H. Mack et al.]{K.-H. Mack$^{1}$\thanks{E-mail:
mack@ira.inaf.it}, M.A. Prieto$^{2}$, G. Brunetti$^{1}$, M. Orienti$^{2,1}$\\
$^{1}$INAF-Istituto di Radioastronomia, Via P. Gobetti 101, I-40129 Bologna, Italy\\
$^{2}$Instituto de Astrof\'{i}sica de Canarias, c/ V\'{i}a L\'actea s/n,
E-38205  La Laguna (Tenerife), Spain
}

\begin{document}
\date{Accepted .... Received ...; in original form ....}
\maketitle
\begin{abstract}
We present new high spatial resolution VLT and VLA observations of a
sample of nine low-power (P$_{1.4 {\rm GHz}} \le 10^{25}$ W/Hz) radio hotspots.
Infrared/optical emission is definitely detected in four of the nine observed 
objects, resulting in a detection rate of at least 45\%. This emission is
interpreted as synchrotron radiation from the electrons accelerated in the hot
spots. 
The integrated spectra of these hotspots reveal typical break frequencies
between
$10^5$ and $10^6$ GHz, two orders of magnitude higher than typically found in 
high-power hotspots. This supports the idea that in low-power hotspots with
their relatively low magnetic field strengths electrons emit most of 
their energy at higher frequencies.
A simple spectral ageing analysis would imply that the emitting electrons have
been injected into the hotspot volume less than $\sim 10^3$ years ago.
We discuss possible scenarios to explain the lack of older electrons in the
hotspot region. In particular, the extended morphology of the NIR/optical
emission would suggest that efficient re-acceleration mechanisms rejuvenate
the electron populations.
 
\end{abstract}

\section{Introduction}
Radio hotspots are  regions of enhanced radio 
emission located at the end of the radio lobes of   powerful radio 
sources. These remote bright regions (usually at several hundred kpc away from 
the galaxy core) mark the ``working surface'' of supersonic jets in which 
their kinetic energy is dissipated into the acceleration of relativistic 
particles. These  relativistic particles  are left behind forming the extended
diffuse radio lobes while the hotspots advance into the ambient medium. As
such, hotspots represent  the end point of an energetic jet: how the energy
of the central radio source is transported, released and maintained in such
remote regions are still open questions. 
However, being relatively isolated regions -- away from the central galaxy
core -- hotspots are ideal laboratories for testing the energetics and the
evolution of a radio galaxy.

\begin{figure*}
\rotatebox{0}{\includegraphics[width=13cm]{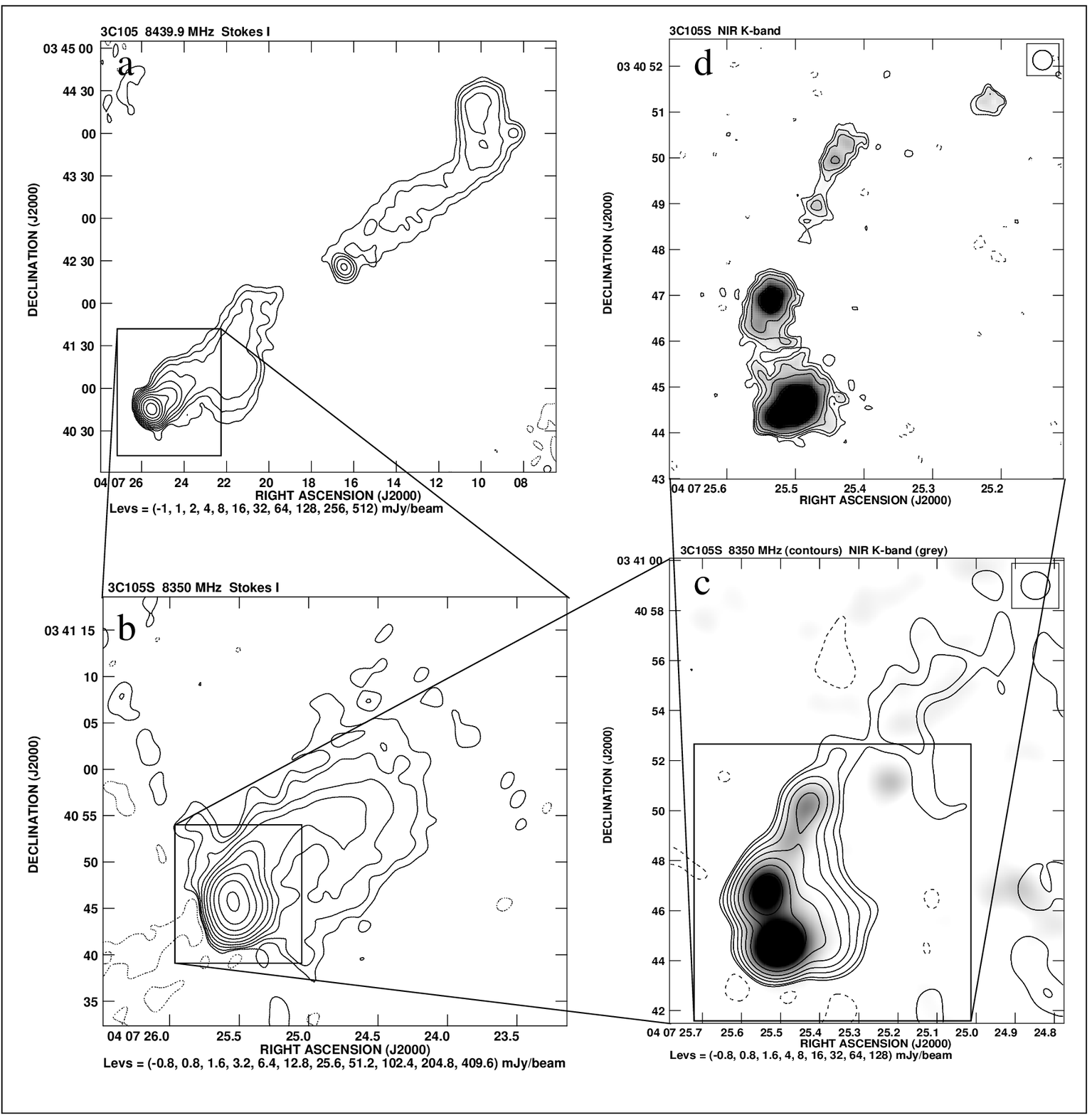}}
\caption{a) 3C\,105: entire source, 8.3 GHz, VLA D-array; b) 3C\,105S:
8.3 GHz, VLA BC-array; c) 3C\,105S: superposition of VLA A-array data 
(contours) and ISAAC K-band image (grey), both smoothed to a common resolution
of $1\farcs1 \times 1\farcs1$; d) 3C\,105S: NIR K-band emission in original 
resolution ($0\farcs43$).}
\end{figure*}

The synchrotron spectral energy distribution between the radio and optical
regime within hotspots, and their radio morphology and polarisation suggest
that the relativistic electrons are accelerated at the shocks that form due to
the impact of the relativistic jets into the ambient medium (e.g. Meisenheimer
2003).
In the context of this scenario the resulting synchrotron spectrum is expected 
to steepen at higher frequencies due to the severe radiative losses that 
electrons experience downstream of the shock region 
(Heavens \& Meisenheimer 1987). 
This is also the reason why optical detections of hotspots are still 
relatively rare although more observations have been performed since the first 
detections at visual wavelengths (e.g., Meisenheimer \& R\"oser 1996; 
Meisenheimer, Yates \& R\"oser 1997; L\"ahteenm\"aki \& Valtaoja 1999; 
Brunetti et al. 2001; Prieto, Brunetti \& Mack 2002; Cheung, Kim \& Song 2005).

At higher frequencies, in the X-ray regime, the search for counterparts of
radio hotspots turned out to be more successful.
Since the initial X-ray detection of the Cygnus A hotspots by Harris, Carilli
\& Perley (1994), many hotspots were found to emit X-rays (e.g., Hardcastle
2003). 
Many of these detections could be explained with synchrotron self-Compton (SSC) 
emission with the magnetic field strength close to the equipartition value. 
A significant number of X-ray emitting hotspots 
however appears to have a larger X-ray intensity than predicted by SSC. 
This excess could be explained by a magnetic field strength much lower than
equipartition, IC emission from
the decelerating jet (Georganopoulos \& Kazanas 2003) or an additional 
synchrotron component (Hardcastle et al. 2004). In particular,
it has been shown by the latter authors that the more luminous hotspots tend 
to show SSC emission, while the weaker radio hotspots emit the extra 
synchrotron component. 
Both X-ray and optical synchrotron emission from hotspots offer the
opportunity to study several of their characteristics which have a number of
fundamental implications for the acceleration of relativistic particles in
general.
They imply either a loss-free transport to the hotspot
(e.g. Kundt 1996), or an in-situ production of energetic particles
with Lorentz factors $>10^5$ (e.g. Meisenheimer et al. 1989).
Transport from the nucleus is however limited by the unavoidable inverse
Compton 
losses experienced by the electrons in the relativistic jet (Gopal-Krishna 
et al. 2001). This makes it difficult to explain the optical emission from hot
spots at sufficiently large distance (i.e.$\geq 150-200$ kpc) from the nucleus
(Brunetti et al. 2003) with this scenario.

In a seminal paper Meisenheimer et al. (1997) discussed the synchrotron spectra 
of hotspots, from radio to optical, through the analysis of a sample of eight 
hotspots. Most of these hotspots have spectra that could in principle be 
accommodated within the standard diffusive shock acceleration model (Bell 1978)
on the basis of the observed radio photon spectral index of 
$\approx$0.5 (S$_\nu \propto \nu^{-\alpha}$). In the context of this scenario 
two other characteristics which describe the spectral shape once the electrons 
have been accelerated are the break frequency, i.e. the maximum energy of the 
oldest electron population in the hotspot volume, driven by the cooling of the 
electrons in the post-shock region, and the cut-off frequency, generated by 
the competition between acceleration and loss mechanisms in the shock region. 
Observationally, the radiation spectral indices typically increase by 
$\Delta \alpha \approx 0.5$ beyond the break frequency, and an exponential 
spectral cut-off develops at the cut-off frequency. 
Brunetti et al. (2003) have shown that the magnetic field strengths are the 
weaker, the higher the break and the cut-off frequencies are, with a scaling
between break frequency and hotspot magnetic field in the form $\nu_b \propto
B^{-3}$ in agreement with theoretical expectations based on the
shock-acceleration model. 
Remarkably, the consequence is that the weaker the 
magnetic field strengths are, the greater the probability is to find hotspots 
with counterparts at optical wavelengths (Blundell et al. 1999).

There are however cases of hotspots that do not fit in the shock-acceleration 
scenario, at least in its simplest form. The extraordinary spectra of some hot 
spots, lacking any indication of synchrotron ageing (e.g., 3C390.3, 
Prieto \& Kotilainen 1997; 3C303, Meisenheimer et al. 1997), and/or showing 
optical extension (e.g., Pictor~A, Thomson, Crane \& Mackay 1995; 3C445, 
Prieto et al. 2002), call for an efficient re-acceleration mechanism 
in situ: basically, the {\it continuous} work of smoothly distributed 
accelerators at different points along the hotspot region.

This paper is the third in a series of articles reporting the results of a 
project aimed at putting studies of spectral emission from hotspots on a 
broader statistical fundament.
In our first paper (Prieto et al. 2002) we showed the optical images
made of the southern hotspot of 3C\,445. These images revealed bright knots 
embedded in diffuse optical emission distributed along the region of the impact 
of the jet into the intergalactic medium, and suggested that continuous 
re-acceleration of electrons (Fermi-II) is at work in this hotspot in addition 
to the prime (Fermi-I) acceleration mechanism.
The second paper (Brunetti et al. 2003) focused on the statistical aspects of
optically detected hotspots identifying the magnetic field
strength in hotspots as the key parameter which determines the probability 
of their detection at optical wavelengths.
The present paper compiles the complete set of optical data from VLT ISAAC and
FORS observations, complemented by high-resolution radio images on a
sample of candidate objects. In Sect. 2 we describe the selection
of our sample and introduce the target sources, Sect. 3 gives the details of
the observations and the data analysis. In Sect. 4 we show the results for the
individual targets. In Sect. 5 we compare the optical magnitudes with the 
corresponding radio data. A summary is presented in Sect. 6. Throughout the 
paper we use the following cosmological parameters: 
${\rm H}_{0}=71\;{\rm km\, s}^{-1} {\rm Mpc}^{-1}$, $\Omega_{\rm \Lambda}= 0.73$, 
$\Omega_{\rm M} = 0.27$.

\begin{figure}
\rotatebox{0}{\includegraphics[width=8cm]{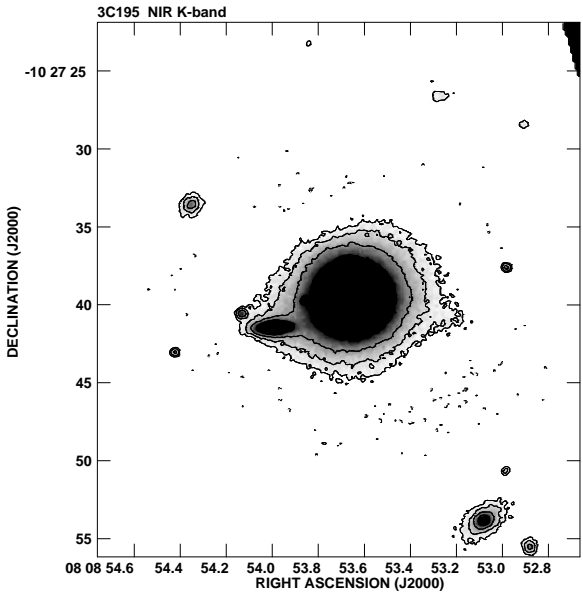}}
\caption{3C195, the host galaxy in K-band. The elongated structure in 
east-west direction discussed by Neff et al. (1995) is a superimposed spiral 
galaxy in combination with a stellar-like object.}
\end{figure}
\begin{figure*}
\rotatebox{0}{\includegraphics[width=13cm]{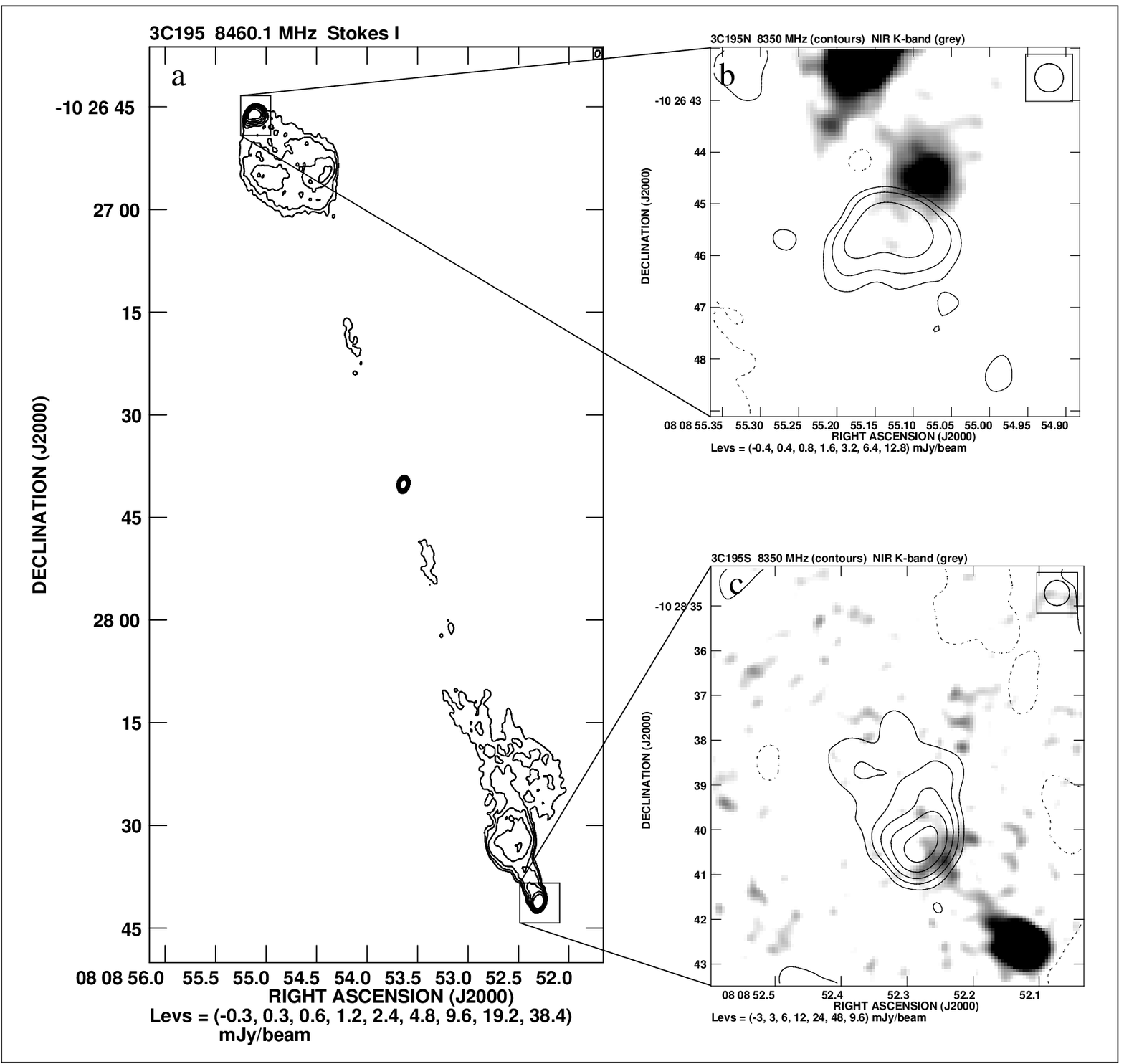}}
\caption{a) 3C\,195: entire source, 8.5 GHz, VLA B-array; b) 3C\,195N:
superposition of VLA A-array data (contours) and ISAAC K-band image (grey),
both smoothed to a common resolution of $0\farcs55 \times 0\farcs55$; 
c) 3C\,195S: superposition of VLA A-array data (contours) and ISAAC K-band
image (grey), both smoothed to a common resolution of $0\farcs55 \times
0\farcs55$.}
\end{figure*}
\begin{figure*}
\rotatebox{0}{\includegraphics[width=13cm]{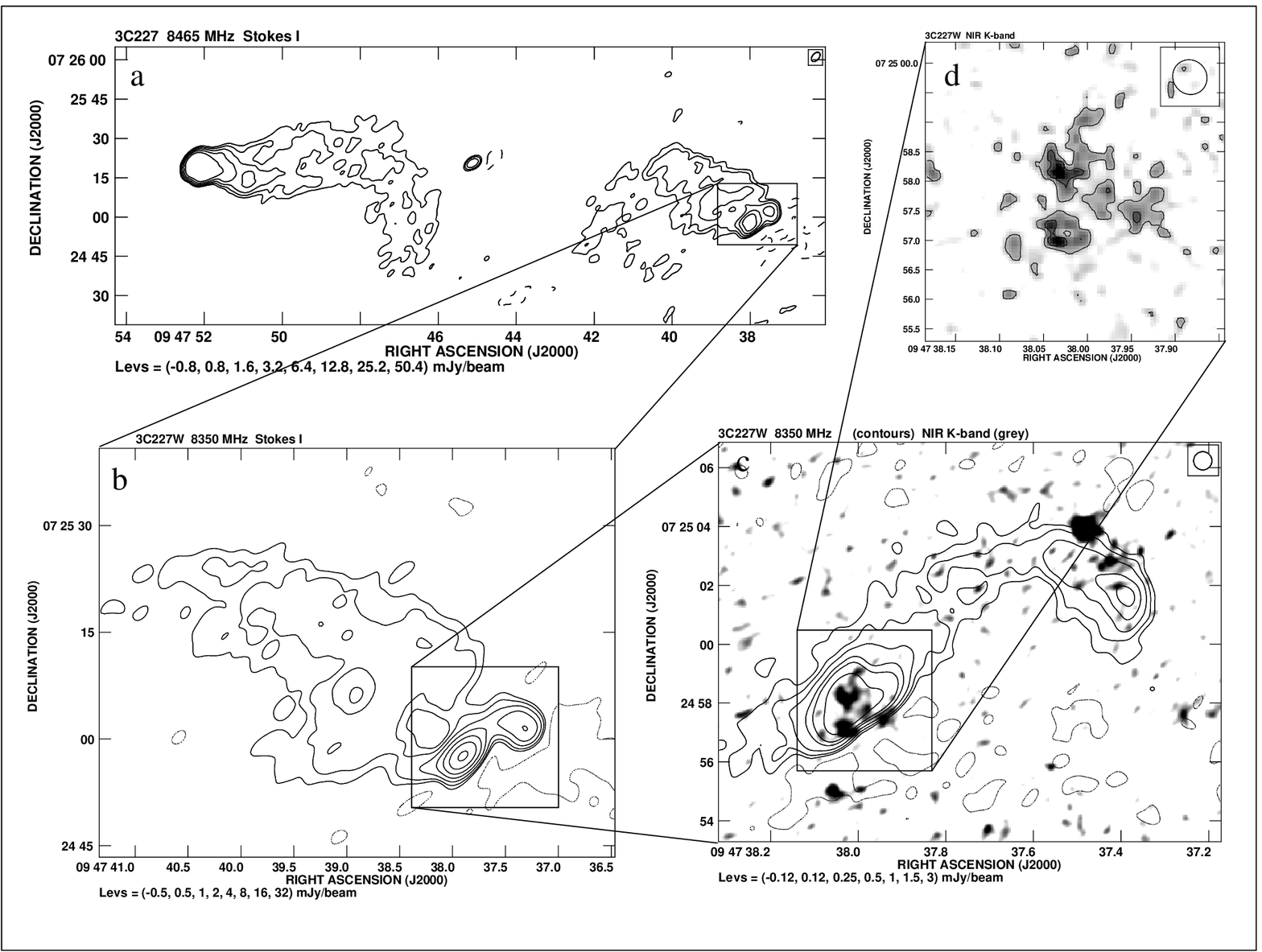}}
\caption{a) 3C\,227: entire source, 8.5 GHz, VLA BC-array; b) 3C\,227W:
8.4 GHz, VLA B-array; c) 3C\,227W: superposition of VLA A-array data (contours)
and ISAAC K-band image (grey), both smoothed to a common resolution of
$0\farcs62 \times 0\farcs62$; d) 3C\,227W: NIR K-band emission in original 
resolution ($0\farcs58$).}
\end{figure*}
\begin{figure*}
\rotatebox{0}{\includegraphics[width=13cm]{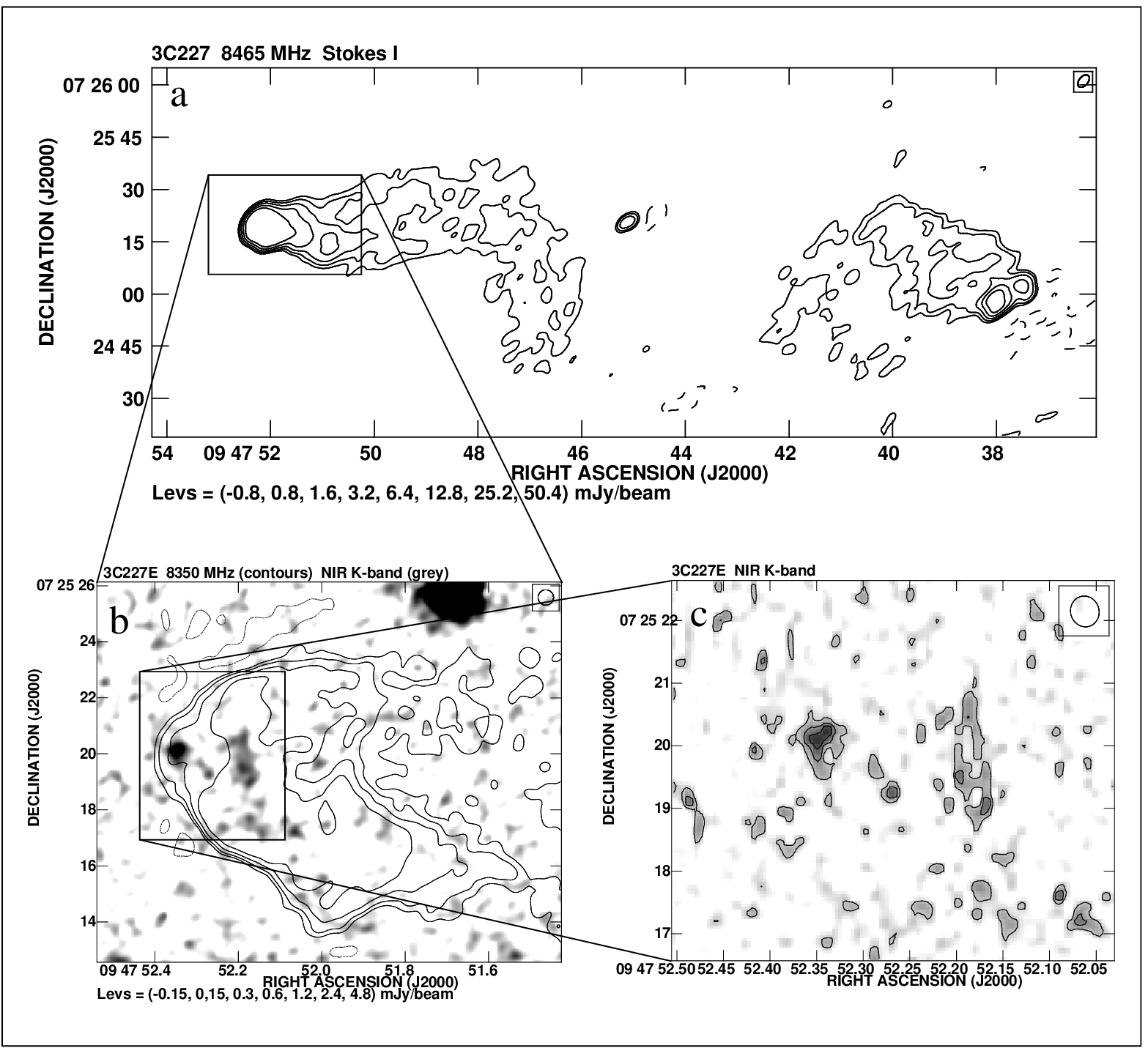}}
\caption{a) 3C\,227: entire source, 8.5 GHz, VLA BC-array; b) 3C\,227E:
superposition of VLA A-array data (contours) and ISAAC K-band image (grey),
both smoothed to a common resolution of $0\farcs55 \times 0\farcs55$;
c) 3C\,227E: NIR K-band emission in original resolution ($0\farcs47$).}
\end{figure*}

\section{Selection of the Sample}

We constructed a sample of hotspots that could be sufficiently
bright to be detected at the VLT. The source list is selected from the complete
sample of radio galaxies presented by Tadhunter et al. (1993) which contains 54
sources with measured redshifts $z < 0.7$ and $-25^{\circ} < \delta < 10^{\circ}$
taken from the 
Wall \& Peacock (1985) 2.7-GHz sample. All these sources have been visually 
inspected using high-resolution maps published by Morganti, Killeen \&
Tadhunter (1993), Morganti et al. (1999), Kapahi et al. (1998), Leahy et al.
(1997) and Hardcastle et al. (1998).
In view of the expected anticorrelation of detection probability and magnetic
field strength (Blundell et al. 1999; Brunetti et al. 2003) we have a priori
excluded the brightest and most compact radio hotspots which emit
most of their energy at low frequencies (radio/mm-band). 
Then we adopted a basic synchrotron model for each radio hotspot by assuming 
a power law spectrum with injection spectral index $\alpha =0.6$ and
a radiative break, 
with $\Delta \alpha =0.5$, at a frequency between the radio and the NIR/optical 
band (e.g. Meisenheimer et al. 1997). We finally used the normalization of
the synchrotron spectrum given by
the radio data for each hotspot and assumed a cut-off frequency of their
synchrotron spectrum at $\nu = 3\cdot 10^{14}$ Hz. A final sample of 10 radio
hotspots turned out to be sufficiently bright to favour a
significant detection in the NIR with VLT.
The selected objects are up to 3 orders of magnitude less luminous
(Tab. 4) than the bright and compact hotspots in Cygnus A
(P$_{\rm 8.3 GHz} \sim 10^{26} {\rm W/Hz}$).

For all of them, high resolution (up to $0\farcs2$ FWHM) radio maps were
obtained by us or collected from the literature and are used for comparison.
In the following we also present maps
at lower resolution to show the large-scale morphological context of the more
detailed high-resolution maps of the hotspots. We also refer the reader to
the images presented by Black et al. (1992) [for 3C\,227 and 3C\,403], 
Leahy et al. (1997) [for 3C\,105, 3C\,327, 3C\,445] and Neff, Roberts \&
Hutchings (1995) [for 3C\,195] for further information on the overal structure
of the sources. 

\subsection{The target sample - Radio properties}
\subsubsection{3C\,105}
3C\,105 at redshift $z=0.089$ is hosted by a Narrow Line Radio Galaxy
(Tadhunter et al. 1993).
The most striking
properties of this source are the protusions in northern and southern direction
of the northern and southern lobes, respectively. These could be interpreted as
remainders of a previous jet direction. Both hotspots are well-defined
with the southern one much brighter than its northern counterpart (see also
Fig. 7 in Leahy et al. 1997).
For our near-infrared detection experiment we have selected the southern
hotspot complex (Fig. 1). Leahy et al. (1997) interpret the two northern
components of this complex morphology as jet emission.
In their scenario, the true jet termination hotspot would be the knot at
$\alpha=04^{\rm h}07^{\rm m}25\farcs6$ and $\delta=03^{\circ}40'46''$,
while the southernmost knot is supposed to be formed by a recent
disconnection event (Cox et al. 1991), i.e. through the impact of a spinning
jet on ambient material of the cocoon wall, subsequent deceleration and 
formation of a new shock front. 
The rich substructure of this hotspot is seen as an indication of a rapidly 
evolving, transient system, presumably influenced by the superposition of a
collimated backflow structure (Leahy et al. 1997).

\subsubsection{3C\,195}
With a redshift of  z$=0.11$ the Narrow-Line Radio Galaxy (Morganti et al.
1997) 3C\,195 is the most distant source in our sample. Neff et al. (1995)
describe it as a large double-lobe source with bright hotspots and a complex
lobe structure, suggesting small changes of the radio axis. The optical
protrusion, or jet, orthogonal to the radio axis, noted by these authors, turns
out to be an underlying spiral galaxy as can be seen at higher resolution
(Fig. 2).
Both, southern and northern hotspot, were selected as targets for the VLT
observations (Fig. 3). 

\subsubsection{3C\,227}
3C\,227 is a Broad-Line Galaxy (Simpson et al. 1996) at
redshift $z=0.0861$. Black et al. (1992) noted that there is no obvious jet
detection in either lobe. 
The hotspots in both lobes have a complex morphology. The western one
consists of two distinct knots (Fig. 4), the eastern one of those is again 
composed of two knots which remind of a primary -- secondary hotspot scenario,
similar to the situation in 3C\,105S. 
Lonsdale \& Barthel (1998) have discussed possible scenarios for such double
hotspots which can be created either via a beam jitter model (as proposed
for 3C\,105S) or a beam deflection model in which the end point of the jet is 
not marked by the primary hotspot but the secondary, which is still being
fed by the deflected, decelerated beam.
On the contrary, 3C\,227E shows a
more filamentary structure (Fig. 5), with an extended area of maximum emission.
Both hotspots were targets of our optical observations.

\subsubsection{3C\,327}
3C\,327 is a Narrow-Line Radio Galaxy (Simpson et al. 1996)
at redshift $z=0.1039$ (Fig. 6a).
The eastern hotspot complex consists of a primary knot and a more diffuse
secondary one north of it. Both features are embedded in a region of diffuse
extended emission (Fig. 6b). The dominant feature in the western hotspot is
the 
slightly curved structure at the very end of the hotspot complex. Leahy et 
al. (1997) mention the possibility that the brighter structure in the south
is part of the radio jet (Fig. 6c).
We performed optical observations of both hotspots.

\subsubsection{3C\,403}
The Narrow Line Radio Galaxy 3C\,403 at redshift 0.059 belongs to the peculiar
class of X-shaped sources which combine two pairs of lobes extending in
different directions from the same core.
One of these lobe systems is typically more extended and diffuse, while the 
other one is shorter, brighter and clearly confined.
The more extended lobes would be the remainders of an 
earlier phase of jet activity, while the shorter, brighter systems would mark
the current working direction of the jet. Black et al. (1992) show the overall
morphology (their Fig. 13). 
Hardcastle et al. (1998) suggest that the knot F6 is the primary hotspot in 
the western lobe.
For our detection experiment we focused on the outermost western hotspot (Fig.
7) at the edge of the younger lobe system which is described as an unusually
uniform hotspot without apparent substructure by Black et al. (1992).

\subsubsection{3C\,445}
The host galaxy of 3C\,445 is a Broad-Line Radio Galaxy (Eracleous \& Halpern
1994) at a redshift of 0.05623. This source has been defined as an example of
recurrent activity sources.
In addition to the bright hotspots in a distance of 315 kpc (north) and 275
kpc (south) of the core, two additional internal hotspots at 47 kpc (north)
and 77 kpc (south), well aligned with the outer hotspots and the nucleus are
found. These are the characteristics of a Double-Double Radio Galaxy (Lara
et al. 1999, Schoenmakers et al. 2000). Both outer hotspots were targets of
our study (Figs. 8 \& 9).  

\section{Observations and data reduction}

\subsection{Near Infrared observations with ISAAC}

The sample of hotspots listed in Table 1 were observed in the
near-IR with ISAAC at the VLT in the J-, H- and K-bands. An exception
were the hotspots 3C\,227E and W and 3C\,327W, which were
observed in K-band only as these are the faintest targets in the
sample. 

The pixel scale of ISAAC in those bands is $0\farcs148$/pix. All observations 
were done in service mode to garantee optimal weather conditions. In each band 
the achieved spatial resolution was exceptional, with a FWHM $< 0\farcs6$ on
average. Observations were collected following a random dithering
pattern about the science target position. Co-addition of the individual frames
was done by applying the standard shift-and-add technique. The total integration
time accumulated on the target was on average about 45 minutes per band. 
In Table 1 a compilation of the ISAAC observations is presented. New data which 
were not used in the first two papers are marked with an asterisk in Col. 5.

\begin{table}
\caption{ISAAC observations of the candidate hotspots. Records with an
asterisk at the observing date mark new data, which were not
used in the first two papers.}
\begin{tabular}{|lcccl|}
\hline
source  & band  & FWHM ($''$)& Magn.& observing date \\
\hline
3C\,105S & K  & 0.43&$ 17.8  \pm 0.2   $ & Aug 20 2001  \\
3C\,105S & H  & 0.46&$ 18.3  \pm 0.1   $ & Sep 24 2002  \\
3C\,105S & J  & 0.70&$ 20.2  \pm 0.2   $ & Aug 23 2002  \\
3C\,195N & K  & 0.40& $>$20.5            & Apr 09 2001  \\
3C\,195N & H  & 0.34& $>$21.6            & Jan 21 2003* \\
3C\,195N & J  & 0.44& $>$22.9            & Jan 21 2003* \\
3C\,195S & K  & 0.40&$  21.3 \pm 0.2   $ & Apr 09 2001  \\
3C\,195S & H  & 0.34& $>$21.6            & Jan 21 2003* \\
3C\,195S & J  & 0.44& $>$22.9            & Jan 21 2003* \\
3C\,227E & K  & 0.47&$  20.6 \pm 0.3   $ & Apr 18 2002  \\
3C\,227WE& K  & 0.58&$  19.8 \pm 0.3   $ & Apr 18 2002  \\
3C\,327E & K  & 0.37&  --                & May 11 2001  \\
3C\,327W & K  & 0.41& $>$22.1            & Apr 18 2001  \\
3C\,403W & K  & 0.53& $>$22.8            & May 11 2003* \\
3C\,403W & H  & 0.56& $>$22.3            & Jun 06 2001  \\
3C\,403W & J  & 0.62& $>$23.6            & Jun 08 2001  \\
3C\,445N & K  & 0.31&$ 19.9 \pm 0.3    $ & Jun 06 2001  \\
3C\,445N & H  & 0.38& $>$21.4            & Sep 18 2001  \\
3C\,445N & J  & 0.46&$ 22.1 \pm 0.3    $ & Sep 18 2001  \\
3C\,445S & K  & 0.75&$ 19.0 \pm 0.1    $ & Jun 07 2001  \\
3C\,445S & H  & 0.59&$ 19.6 \pm 0.2    $ & Jul 06 2001  \\
3C\,445S & J  & 0.49&$ 20.2 \pm 0.1    $ & Jun 23 2001  \\\hline
\end{tabular}
\end{table}
\subsection{Optical observations with FORS}

Follow-up observations of the hotspot sample were carried out with FORS at the
VLT in the R and B bands, with the exceptions of 3C\,327W and E, 3C\,403W, and 
3C\,445N. 3C\,327E is confused by a background galaxy,
3C\,327W is too faint for optical observations and was not detected in the
NIR; 3C\,403W was not detected in the NIR either and the follow-up optical
observations were done in I-band only. Only 3C\,445S was in addition observed 
in the I and U bands. The pixel scale provided by the FORS configuration used
with
these bands is $0\farcs20$/pix. The observation technique used was the same as
that described above with ISAAC. The total integration time is about 1 h
in the I- and B-bands, 0.5 h in the R-band, and 1.6 h in the U-band. The spatial
resolution achieved in these observations was on average FWHM $\sim 0\farcs6$. 
In Table 2 a compilation of the FORS observations is presented. New data which 
were not used in the first two papers are marked with an asterisk in Col. 5.

\begin{table}
\begin{center}
\caption{FORS observations of the candidate hotspots. Records with an
asterisk at the observing date mark new data, which were not
used in the first two papers.}
\begin{tabular}{|lcccl|}
\hline
source & band  & FWHM ($''$) & Magn.&observing date \\
\hline
3C\,105S  & R   & 0.6  &$  22.8 \pm 0.1 $ & Nov 26 2003* \\
3C\,105S  & B   & 0.74 &$  24.5 \pm 0.3 $ & Nov 26 2003* \\
3C\,195N  & R   & 0.68 &  $>25.8$         & Dec 18 2003* \\
3C\,195N  & B   & 0.68 &  $>25.7$         & Nov 30 2003* \\
3C\,195S  & R   & 0.68 &$  25.4 \pm 0.2 $ & Dec 18 2003* \\
3C\,195S  & B   & 0.68 &  $>25.7$         & Nov 30 2003* \\
3C\,227WE & R   &  0.8 &$  24.4 \pm 0.2$  & Dec 18 2003* \\
3C\,227WE & B   &  0.8 &  $25.4 \pm0.2$   & Dec 18 2003* \\
3C\,227E  & R   &  0.8 &$  23.9 \pm 0.2$  & Dec 18 2003* \\
3C\,227E  & B   &  0.8 &$  25.6 \pm 0.2$  & Dec 18 2003* \\
3C\,403W  & I   & 0.65 & $>$24.0          & Sep 14 2001  \\
3C\,445S  & I   & 0.66 &$ 21.5\pm 0.1$    & Sep 15 2001  \\
3C\,445S  & R   & 0.58 &$ 22.0\pm 0.2$    & Nov 02 2002  \\
3C\,445S  & B   & 0.56 &$ 22.9\pm 0.2$    & Nov 02 2002  \\
3C\,445S  & U   & 0.72 &$ 23.2\pm 0.2$    & Jul 03 2003* \\\hline
\end{tabular}
\end{center}
\label{tab_fors}
\end{table}

\begin{figure*}
\rotatebox{0}{\includegraphics[width=13cm]{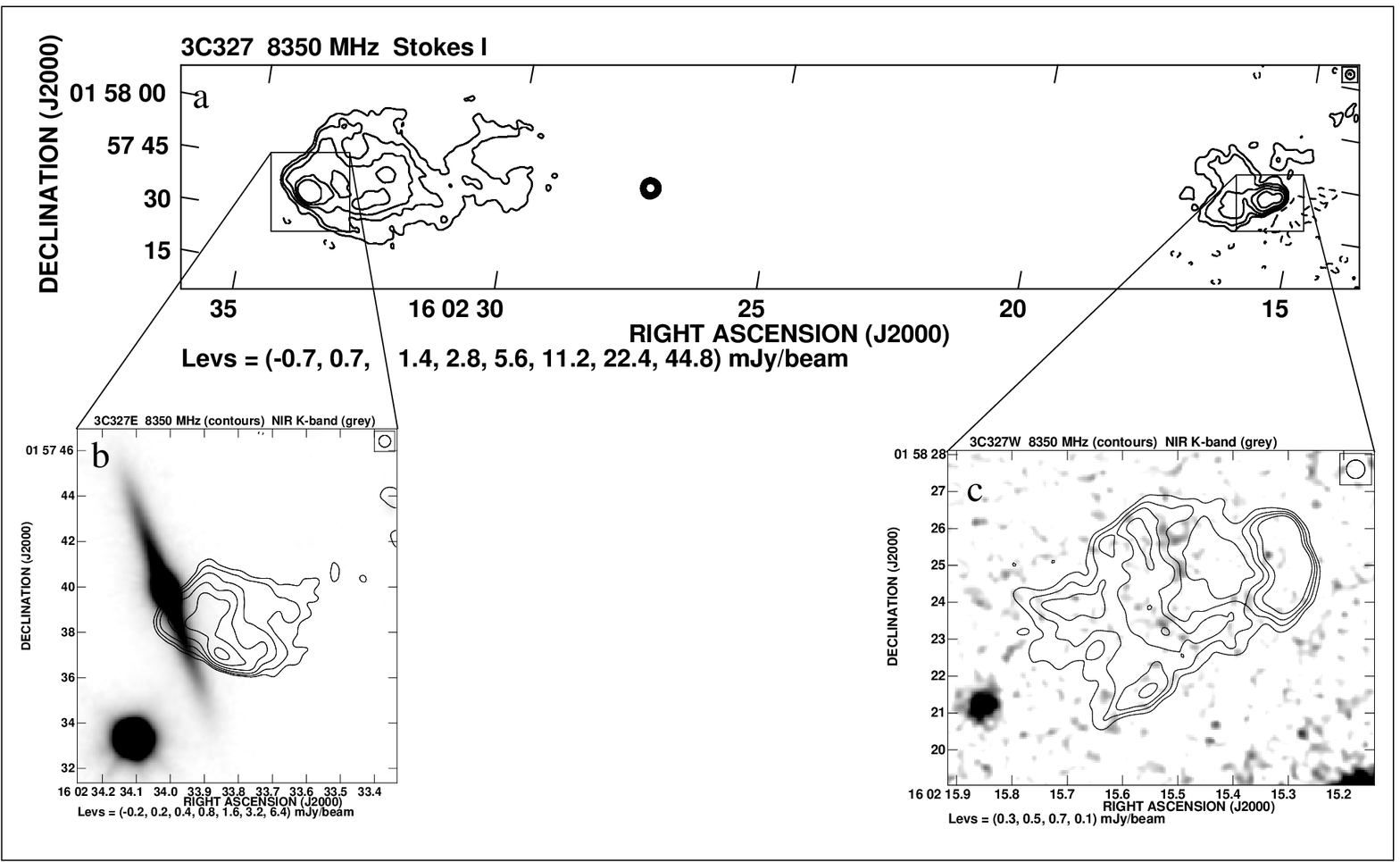}}
\caption{a) 3C\,327: entire source, 8.4 GHz, VLA  BC-array; b) 3C\,327E:
superposition of VLA A-array data (contours) and ISAAC K-band image (grey),
both smoothed to a common resolution of $0\farcs52 \times 0\farcs52$; 
c) 3C\,327W: superposition of VLA A-array data (contours) and ISAAC K-band
image (grey), both smoothed to a common resolution of $0\farcs50 \times 
0\farcs50$.}
\end{figure*}

\subsection{Radio data}

Between April 2002 and January 2004, we observed all targets, with the
exception of 3C\,227 and 3C\,327, with the VLA at 1.4, 5.0 and/or 8.4 GHz
either in A or B configuration, in order to construct multi-frequency radio
continuum spectra of the hotspots. For the
source 3C\,195 we analysed further archival data at 5.0 and 15.0 GHz.  
Information on the observing runs are provided in Table \ref{radio_obs}. 
The data reduction was carried out following the standard procedures
for the VLA implemented in the NRAO AIPS package. \\
For each observing run about 3-5 min, depending on the frequency, were spent on
the primary calibrator 3C\,286. Secondary calibrators, chosen to minimize the
telescope slewing time were observed for 1.5 min every 5 min. The flux
densities were calibrated according to the scale of Baars et al. (1978). At 1.4
GHz it was necessary to image a few confusing sources falling into the primary
beam in order to have an accurate measurement of the flux density. 
The final images were produced after a few phase-only self-calibration
iterations. \\
In addition, these data were complemented by archive data from the NRAO VLA 
archive, observed as part of the project AB534, which were originally
published by Black et al. (1992) and Leahy et al. (1997). 
In Table 3 a compilation of the VLA observations is presented. New data which 
were not used in the first two papers are marked with an asterisk in Col. 5.

\begin{table}
\begin{center}
\tabcolsep0.15cm
\caption{VLA observations of the candidate hotspots}
\begin{tabular}{|lccccl|}
\hline
source & frequency& conf.& beamsize &beam p.a. &obs. date \\
       & [GHz] &    & [$''\times\; ''$]& [$^{\circ}$] &       \\\hline
&&&&&\\
3C\,105S    &  1.4  & A & 1.23$\times$1.18 &  35 & Jul 19 2003*\\
3C\,105S    &  4.8  & A & 0.37$\times$0.35 &  78 & Jul 19 2003*\\
3C\,105S    &  8.4  & A & 0.54$\times$0.26 & $-$71 & May 25 1990\\
3C\,195N, S &  1.4  & B & 6.22$\times$4.01 & $-$20 & Jan 17 2004*\\
3C\,195N, S &  4.9  & B & 1.98$\times$1.19 &  31 & Aug 18 1986\\
3C\,195N, S &  8.4  & B & 1.04$\times$0.67 & $-$15 & Jan 17 2004*\\
3C\,195N, S &  8.4  & A & 0.32$\times$0.21 &  12 & Aug 03 1991\\
3C\,195N, S & 15.0  & B & 0.50$\times$0.35 &   3 & Sep 02 2006*\\
3C\,227W, E &  8.4  & A & 0.38$\times$0.22 &  50 & May 25 1990\\
3C\,327W    &  8.4  & A & 0.35$\times$0.22 &  48 & May 25 1990\\
3C\,403W    &  1.4  & A & 1.82$\times$1.37 &  58 & Apr 18 2002\\
3C\,403W    &  4.8  & A & 0.47$\times$0.35 &  55 & Apr 18 2002\\
3C\,403W    &  8.4  & B & 0,97$\times$0.85 & $-$45 & Apr 07 1989\\
3C\,445N    &  1.4  & A & 1.43$\times$1.04 &  15 & Jul 19 2003*\\
3C\,445N    &  4.8  & A & 0.46$\times$0.37 & $-$23 & Jul 19 2003*\\
3C\,445N    &  8.4  & A & 0.27$\times$0.24 & $-$31 & Apr 18 2002\\ 
3C\,445S    &  1.4  & A & 1.42$\times$1.02 &  11 & Jul 19 2003*\\
3C\,445S    &  4.8  & A & 0.46$\times$0.37 & $-$23 & Jul 19 2003*\\
3C\,445S    &  8.4  & A & 0.26$\times$0.22 &  12 & Apr 18 2002*\\
&&&&&\\
\hline
\end{tabular}
\label{radio_obs}
\end{center}
\end{table}

\begin{table}
\begin{center}
\tabcolsep0.15cm
\caption{VLA flux densities and angular sizes of the candidate hotspots}
\begin{tabular}{|lccccccc|}
\hline
source &S$_{\rm 1.4}$&S$_{\rm 4.8}$&S$_{8.4}$&S$_{15}$&$\theta_{\rm maj}$&
$\theta_{\rm min}$ & P$_{8.4 \rm GHz}$ \\
 &[mJy]&[mJy]&[mJy]&[mJy]&[$''$]&[$''$] & [$10^{23}$ W/Hz]\\
\hline
&&&&&\\
3C\,105S   & 2790  & 1268 & 873 & $-$  & 7.0 & 5.0 & 169\\
3C\,195N   &  250  &   90 &  45 & 22 & 1.7 & 1.4   &  13.7\\
3C\,195S   &  440  &  140 &  67 & 34 & 1.8 & 1.1   &  20.3\\
3C\,227E   &  $-$    & $-$    &  47 & $-$  & 1.7 & 1.4 & 8.5\\
3C\,227WE  &  $-$    & $-$    &  37 & $-$  & 1.9 & 1.2 & 3.8\\
3C\,327W   &  $-$    & $-$    &  21 & $-$  & 2.4 & 1.0 & 5.6\\
3C\,403W   &   94  &   50 &  33 & $-$  & 5.4 & 4.1 & 4.7\\
3C\,445N   &  160  &   58 &  35 & $-$  & 3.5 & 2.0 & 2.6\\
3C\,445S   &  520  &  135 &  81 & $-$  & 6.0 & 3.0 & 6.0\\ 
&&&&&\\
\hline
\end{tabular}
\label{radio_flux}
\end{center}
\end{table}

\begin{figure*}
\rotatebox{0}{\includegraphics[width=13cm]{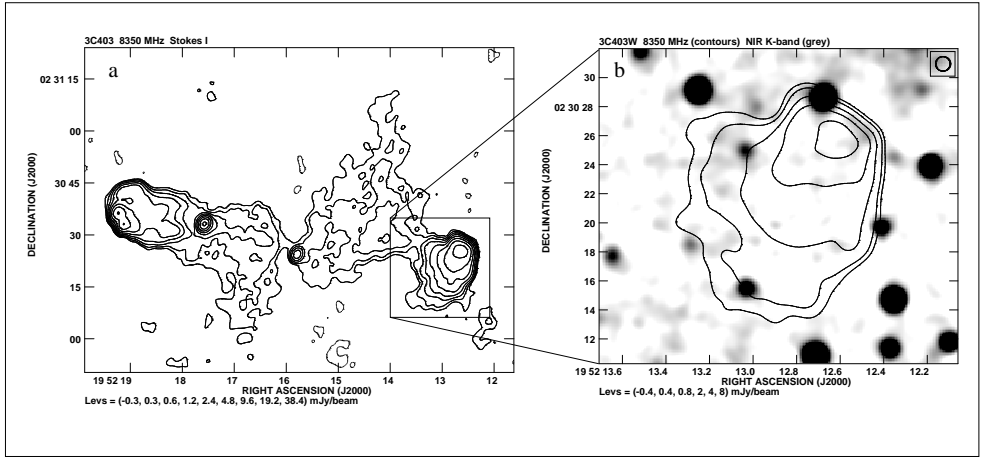}}
\caption{3C\,403: entire source, 8.4 GHz, VLA B-array; b) 3C\,403W: 
superposition of VLA B-array data (contours) and ISAAC K-band image (grey), 
smoothed to a common resolution of $1\farcs0 \times 1\farcs0$.}
\end{figure*}

\begin{figure*}
\rotatebox{0}{\includegraphics[width=13cm]{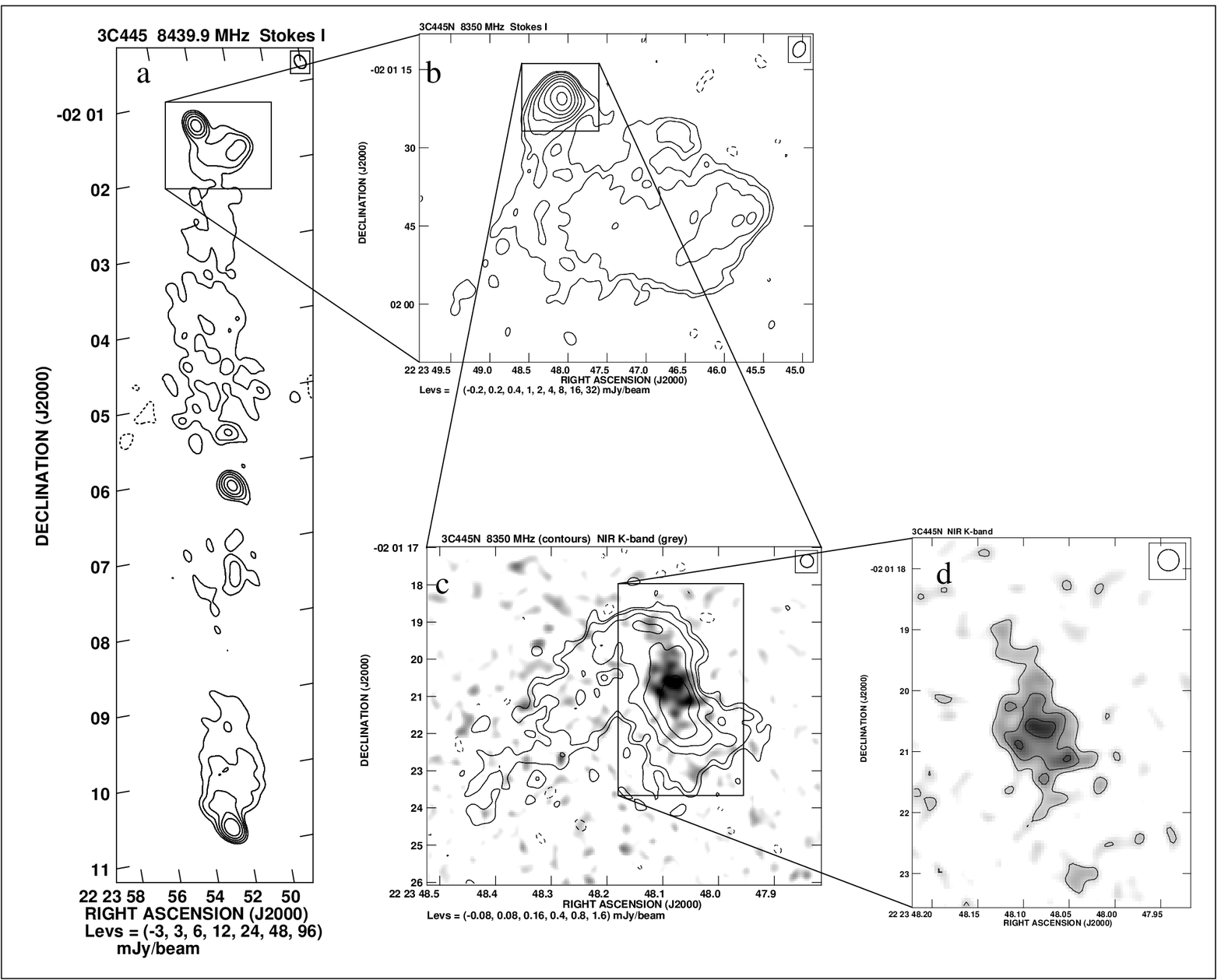}}
\caption{3C\,445: entire source, 8.4 GHz, VLA C-array; b) 3C\,445N:
8.4 GHz, VLA B-array; c) superposition of VLA A-array data (contours) and 
ISAAC K-band image (grey), both smoothed to a common resolution of
$0\farcs35 \times 0\farcs35$; d) 3C\,445N: NIR K-band emission in original
resolution ($0\farcs31$).}
\end{figure*}
\begin{figure*}
\rotatebox{0}{\includegraphics[width=13cm]{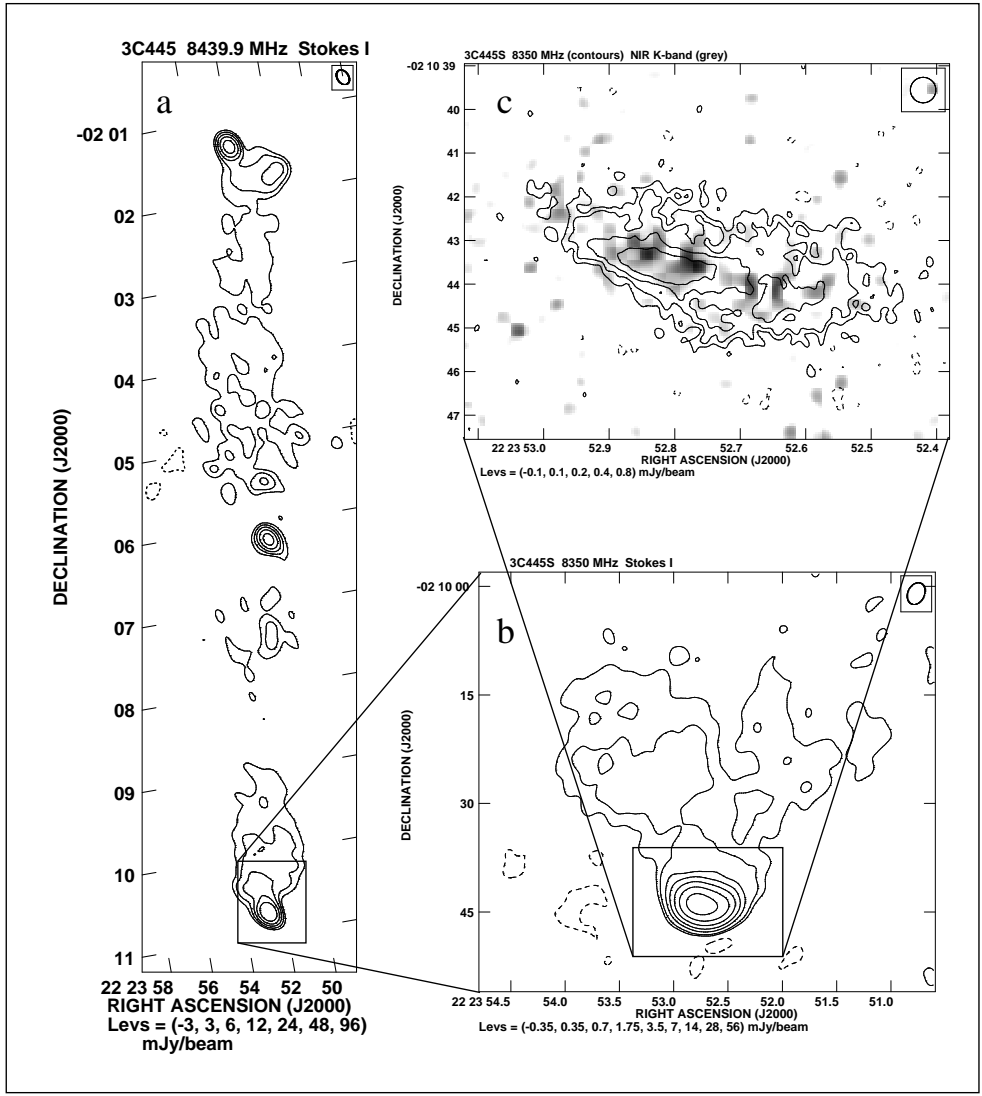}}
\caption{3C\,445: entire source, 8.4 GHz, VLA C-array; b) 3C\,445S: 8.4 GHz,
VLA B-array; c) superposition of 
VLA A-array data (contours) and ISAAC K-band image (grey), both smoothed to a 
common resolution of $0\farcs58 \times 0\farcs58$. }
\end{figure*}

\begin{figure}
\includegraphics[width=8.7cm]{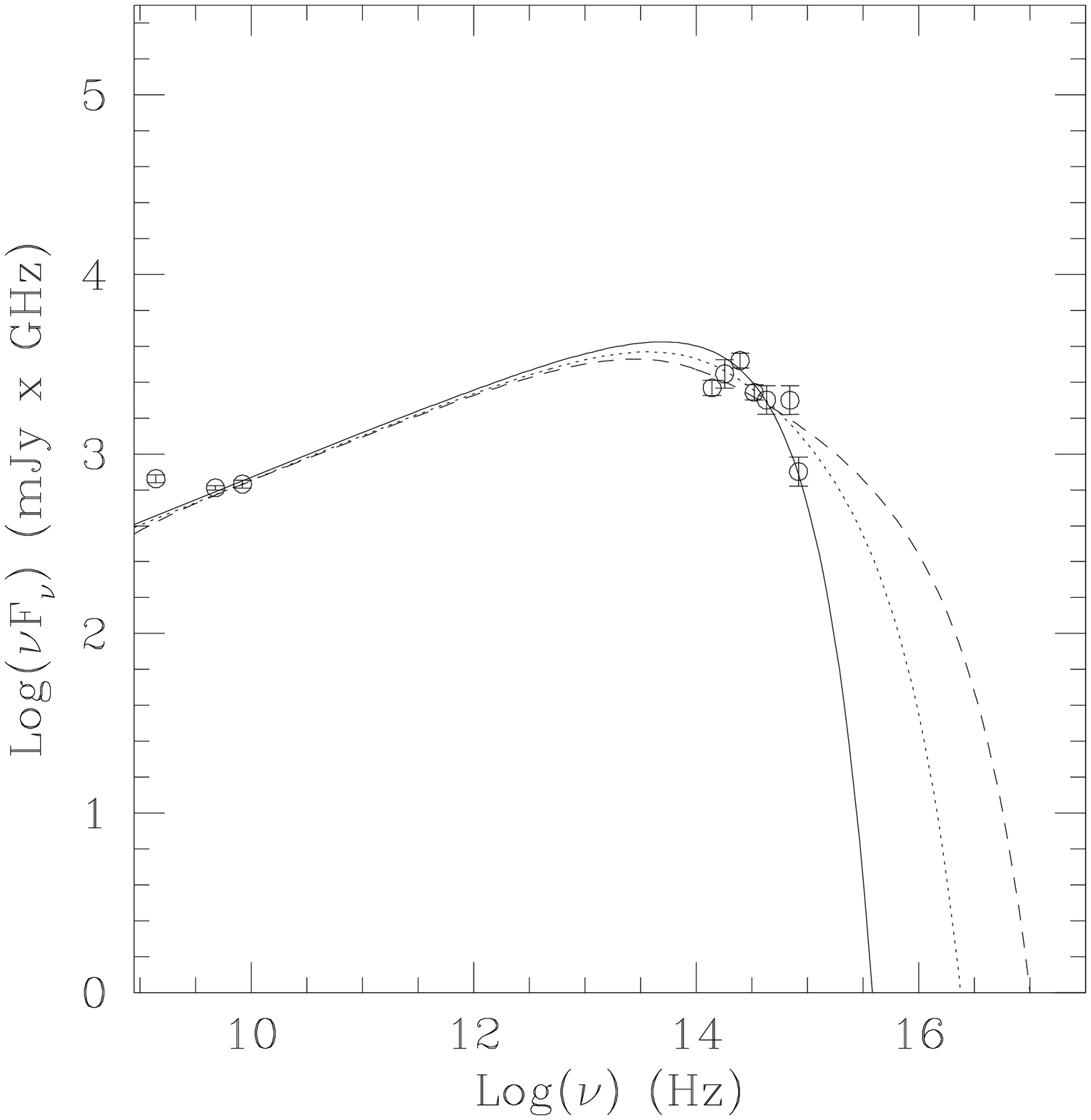}
\caption{Examples of model spectral energy distributions of the hotspot of
3C\,445S for three different break frequencies 
(assuming $\alpha_{\rm inj} = 0.75$) to illustrate a comparison between 
model expectations and data. The models are normalised at the radio flux 
densities. The different break frequencies are $\nu_b = 5.0\cdot10^{14}$ Hz, 
$\nu_c/\nu_b= 1$ (solid line); $\nu_b = 1.2\cdot10^{14}$, $\nu_c/\nu_b=50$ 
(dotted line) and $\nu_b = 7.0\cdot10^{13}$ Hz, $\nu_c/\nu_b=400$ (dashed line).
}
\end{figure}

\section{Results}
To identify a possible NIR/optical counterpart of the radio hotspots
we superimposed the radio images with the NIR/optical data. To
this purpose, the optical images were brought on the same grid and 
orientation as the radio images. Several sources of the USNO-B1.0 survey 
(Monet et al. 2003) were
identified in the images and used as reference sources for a transformation
of the coordinate systems using AIPS task XTRAN.
The alignment between radio and optical images was cross-checked by
superposition of point-like radio and optically emitting regions like the
host galaxies (when covered by the optical image) or other AGN in the field.
We estimate the typical remaining uncertainties to be of the order of 
$0\farcs4$.
In order to assure the correct determination of flux densities at the various
wavebands in the same regions, all images were convolved to a common resolution
using AIPS task CONVL.
The hotspot flux densities and angular sizes were measured in the
images using the AIPS task JMFIT which performs a Gaussian fit, while in
extended structures the flux density was obtained by TVSTAT. Observational
parameters of the hotspots are reported in Tables 1 and 2 for NIR/optical
observations and Table 4 for the radio observations.
The main uncertainties of the flux density measurements come
from amplitude calibration errors, usually between 3\% and 5\%, while the
r.m.s. noise level in the image plane is not relevant.

The optical magnitudes were determined using the IDL-based task ATV
in various iterations to identify the optimum integration area around the
targets and correcting for any remaining local background instabilities, which
however dominate the measurement errors. 
Magnitudes were subsequently transformed into jansky using standard
formulae (Campins, Rieke \& Lebofsky 1985; Allen 1972).

\subsection{Optical detections}

We unambiguously detected optical emission connected with the radio emission 
in four of the nine hotspots
of the final sample (3C\,327E excluded): 3C\,105S, 3C\,227W, 3C\,445N and
3C\,445S. In addition there are two doubtful cases 
(3C\,195S and 3C\,227E), for which the successful detection of an NIR
counterpart cannot be confirmed yet on the basis of the present imaging data.
No detections were found for 3C\,195N, 3C\,327W and 3C\,403W. The finally
reached detection rate is 67\% (decreasing to 45\% if the optical 
counterparts of 3C\,195S and 3C\,227W are not confirmed). Such a 
success rate is much higher than usually found in the literature and may be 
due to the low magnetic field strengths in these, specifically selected, 
hotspots. 

In Figs. 1 and 3 - 9 we present the NIR K-band emission for each of our 
target superimposed to the radio image, both smoothed to the same resolution.
In case of a large discrepency between the resolutions of the convolved
and the original images the optical emission is also shown at its original
resolution.
 
\subsection{Comments on individual sources}
\subsubsection{3C\,105}
The southern hotspot of 3C\,105 is characterized by three well-resolved
asymmetric sub-structures, each one with a clear NIR counterpart
(Fig. 2). The NIR emission at $\alpha = 04^{\rm h}07^{\rm m}25\farcs22$,
$\delta=03^{\circ}40\arcmin51\farcsec2$ does not have a radio counterpart
and is considered to be an unrelated background source.
The northern and southern components are located $\sim 3\farcs8$ (6.2
kpc) and $2\farcs1$ (3.4 kpc) respectively 
from the central component, which is also the brightest one in radio,
while in the optical band the southern component accounts for the higher flux
density. All components have a spectral index between 8.4 GHz and
K-band of $\alpha \sim 1.0$.  
A more detailed analysis of this impressive hotspot complex will be 
presented in a forthcoming paper.

\subsubsection{3C\,195}
The high-resolution image of the northern hotspot of 3C\,195 reveals the
detected NIR emission to be unrelated to the radio emission.
The situation in the southern hotspot is more difficult to judge. The NIR
observations show extended emission within the contours of the radio emission.
However, the shape of the optical emission seems to be unrelated to the radio
morphology and its barycentre is shifted by about one beam size to the
southwest with respect to the radio peak.
This and the relative weakness of the NIR emission make a successful detection
arguable.

\subsubsection{3C\,227} 
The western lobe is characterized by two asymmetric radio hotspots located
about 10$''$ ($\sim$ 16 kpc) apart from each other. Their flux density ratio is
S$_{\rm WE}$/S$_{\rm WW}$ $\sim$ 3.3. 
Only the eastern component has a clear positional counterpart with the NIR
emission, while the situation of the westernmost component is more uncertain 
with the optical counterpart located outside the compact region of the radio 
hotspot (Fig. 4c).
The spectral indices between 8.4 GHz and K-band are $\alpha_{\rm WE} \sim 0.9$ 
and $\alpha_{\rm WW} > 1.0$ for the eastern and the western components (assuming 
that the optical emission is not connected with the western radio component), 
respectively. 
Hardcastle, Croston \& Kraft (2007) found X-ray emission from both the eastern
and western 
components of the western hotspot of 3C 227. Remarkably the radio and X-ray 
peaks are not coincident and these authors discuss possible consequences of the 
positional offsets on the mechanisms responsible for the observed emission. 
Clearly deeper optical observations would greatly help to constrain positional 
offsets between the counterparts of the hotspot at different wavelengths 
and to eventually confirm the detection of the optical counterpart of the 
western component of 3C 227W.

Also the eastern lobe shows underlying NIR emission in the area of its hot 
spot (Fig. 5). These two distinct features are both relatively weak. The 
eastern component (Fig. 5c) at the tip of the lobe is basically point-like,
the elongated feature to the west is fainter and roughly coincides with the
region of maximum radio brightness. In principle, this extended NIR emission
could derive from the radio hotspot, although in order to verify the optical
detection of the hotspot beyond any doubt much deeper optical imaging is
required.
A comparison with the X-ray emission reported by Hardcastle et al. (2007)
remarkably shows coincident positions of the extended NIR emission with a
chain of point-like X-ray emitters (Fig. 4 in Hardcastle et al.).

\subsubsection{3C\,327}
The western hotspot does not show any significant NIR counterpart. This 
results in a lower limit for the spectral index $\alpha >1.0$.
The image of the eastern hotspot is confused by an
underlying spiral galaxy.
As it is therefore impossible to decide whether the hotspot emission reaches 
up to the NIR band, 3C327E has been removed from our sample. 

\subsubsection{3C\,403}
This source at relatively low Galactic latitude is situated in a star-rich
environment. None of the significant NIR detections within the radio contours
seems to be related to the radio hotspot.
Kraft et al. (2005) report extended X-ray emission from the western lobe
of 3C\,403 and suggest inverse Compton scattering of beamed IR/optical photons
from the active nucleus. On the other hand, these authors find several X-ray
and optical emitting knots associated with the eastern jet, which however are
not covered by our data.

\subsubsection{3C\,445}
Both hotspots of 3C\,445 are characterized by extended emission, 
about $2\farcs5\times1\farcs5$ ($\sim$2.7 kpc$\,\times\,$1.6 kpc) for the North
and about $7''\times4''$ ($\sim$7.5 kpc$\,\times\,$4.3 kpc) for the South,
detected also in the optical band. 
The northern one is elongated in the NE-SW direction (Fig. 8c), while the
southern hotspot shows an arc-like structure (Fig. 9c).
Both hotspots have a spectral index between 8.4 GHz and K-band of
$\alpha \sim$ 0.9.
Prieto et al. (2002) underlined the importance of 3C\,445S that is a clear 
example of a hotspot whose properties are difficult to reproduce with a simple 
shock-acceleration model. The optical images of the hotspot region show an 
arc-like structure with bright compact knots, identified as 
{\it local particle accelerators}, embedded into more diffuse optical emission. 
These knots are interpreted as the result of a small-scale dentist drill 
which deposits the kinetic energy of the jet in several locations, while 
the diffuse emission may indicate that in-situ particle re-acceleration, 
due to Fermi-II mechanisms, is at work in the whole hotspot region.

A detailed analysis of these hotspots will be presented in a forthcoming
paper.

\section{Synchrotron spectra}

In low-power radio hotspots the radiative life-times of the electron
populations 
is expected to be longer than in high-power radio hotspots, likely due to lower
values of the magnetic field (Blundell et al. 1999; Brunetti et al. 2003). In
such a situation, the
break frequency may occur at IR and optical wavelengths.

In the context of the shock-acceleration scenario two important frequencies
exist in the synchrotron spectrum of hotspots: the break frequency, $\nu_b$,
that marks the maximum frequency of the synchrotron emission from the oldest 
electrons still confined in the hotspot volume, and the cut-off frequency, 
$\nu_c$, that marks the maximum frequency of the synchrotron radiation emitted 
by the electrons accelerated at the shock 
(e.g. Meisenheimer et al. 1989; Brunetti et al. 2002).
Constraining these frequencies provides fundamental implications on the
efficiency
of the acceleration of particles at the shock and on the physical conditions in
the shock region (e.g. magnetic field strength).
Reliable constraints can be obtained if the optical data points sample the 
spectrum at $\nu > \nu_b$ (e.g., Meisenheimer et al. 1997; Brunetti et al. 2001;
Wilson, Young \& Shopbell 2001), i.e. in the case of hotspots emitting most of
their synchrotron radiation at mm--far-IR wavelengths.
On the other hand, the bulk of the synchrotron radiation from
the hotspots in our sample is emitted at optical wavelengths and thus it is 
possible to constrain only $\nu_b$;
an example is given in Fig. 10, where we show the case of 3C\,445S.

Thus, as $\nu_b/\nu_c$ is poorly constrained we decided to fit the
spectra of the hotspots with a simple single-injection synchrotron model
(Jaffe \& Perola 1973) that is an over-simplification (with $\nu_b = \nu_c$) 
but still provides a viable approach to obtain statistical
measures of $\nu_b$ or upper limits to $\nu_b$ in the case
of optical non-detections. More specifically, for 3C\,445N and 3C\,445S we did
not consider the flux density
at 1.4 GHz because of the strong contamination from the lobe.
The resulting break frequencies for most of the hotspots
range between 10$^{5}$ and 10$^{6}$ GHz.
The upper limit of the break frequency obtained for hotspots without optical
counterparts is $\leq$ 5$\times$10$^{4}$ GHz.
In the case of 3C\,327W the availability of only two data points
did not allow us to perform any synchrotron model on their spectrum. 
Model fits to the spectra of the hotspots are shown in
Fig. 11, while fit parameters are reported in Table 5.  

From the break frequency $\nu_{\rm br}$ [GHz] it is possible to determine the
radiative age $t_{\rm rad}$ [Myr] of the electron populations, once the
magnetic field strength $B$ [$\mu$G] is known, by:

\begin{equation}
t_{\rm rad} \simeq 1610\; B^{-3/2} \frac{1}{\sqrt{\nu_{\rm br} (1+z)}}
\label{eq_trad}
\end{equation}

where the magnetic field equivalent to the cosmic microwave
background has been neglected being much smaller than the typical
magnetic fields in hotspots.
These fields were computed assuming equipartition conditions
(Table 5), following the approach of Brunetti, Setti \& Comastri (1997)
assuming a minimum Lorentz factor of the relativistic electrons 
$\gamma_{\rm min} = 100$, equal proton and electron energy densities,
and an ellipsoidal geometry with a filling factor of unity 
(i.e. the hotspot volume is fully and homogeneously filled by 
relativistic plasma). 

If we consider the break frequency inferred from the fits we find that
the radiative ages of the electron populations with detected optical
emission located in such magnetic fields are of the order of  $10^{3}$ years
(Table \ref{tab_synchro}).
These relatively short time scales imply either that the injection of
relativistic plasma from the jet is not continuous but happens on a period
of $\approx 10^3$ years, or that adiabatic expansion in the hotspot takes
place in $\leq 10^3$ years and the emission from older
electrons is thus just faded away.

In the second scenario, however, adiabatic expansion is expected to affect
the properties of the emitting plasma in a time scale that should be
significantly larger than a crossing time-scale of the hotspot 
at the speed of sound (or Alfv\'en speed), $l_{HS}/c_s > l_{HS}/c$. 
The crossing time scale should be smaller than (or comparable to) the ages
measured in our hotspots yielding the following condition:

\begin{equation}
\left( {{B}\over{B_{eq}}} \right)^{3/2} {{l_{HS}}\over{\rm kpc}}
< 1
\end{equation}

\noindent
that would require either that the magnetic fields in the hotspots 
are substantially smaller than the equipartition field, or that compact
(sub-kpc), over-pressured, emitting regions, are embedded in the hotspot.
If these two conditions are not realised the adiabatic losses
are not fast enough to act within a time-scale of $\approx 10^3$ years
and the absence of older electrons in the hotspots is not easy to
account for in the context of a scenario of continuous jet activity.
On the other hand, the radiative age of the hotspots measured from the 
synchrotron break frequency is based on the assumption that electrons, 
accelerated at the shock, cool in the post shock region in absence 
of any additional (efficient) re-acceleration mechanism.
This might not be the case since the evidence for diffuse synchrotron
emission on kpc scale in several hotspots in our sample 
suggests that in situ (additional) particle re-acceleration mechanisms may 
be at work in the hotspot volume. 
In this case the radiative ages estimated from the break frequency 
should represent a lower limit to the real age of the electrons.
For instance, Prieto et al. (2002) suggested that Fermi II mechanisms with 
a typical electron re-acceleration time-scale $\approx 10^4$ years are required
to explain the diffuse optical emission in the hotspot 3C 445S. 
If these mechanisms are slightly more efficient than previously thought they
may 
contribute to balance the radiative losses of the emitting particles, in 
which case electrons injected in the hotspot volume more than $10^3$ years 
ago may still be emitting at optical frequencies.

\begin{table}
\begin{center}
\caption{Physical parameters}
\tabcolsep1mm
\begin{tabular}{|lrrrrr|}
\hline
Source&$\nu_{\rm br}$&$\alpha$&B$_{\rm eq}$&t$_{\rm rad}$&distance \\
 &[GHz]& &[$\mu$G]&[10$^{3}$ yr]& [kpc] \\
\hline
&&&&\\
3C\,105S  & $1.37   \cdot$10$^{5}$ &0.75 & 75 &  6.4 & 278  \\ 
3C\,195N  & $<2.7   \cdot$10$^{5}$ &0.95 & 62 &$>$6.0& 117  \\
3C\,195S  & $5.34   \cdot$10$^{5}$ &1.00 & 78 &  3.0 & 127  \\
3C\,227WE & $3.0    \cdot10^5$     &0.65 &126 &  2.0 & 173  \\
3C\,227E  & $1.14   \cdot10^6$     &0.75 & 99 &  1.5 & 169  \\
3C\,403W  & $<$2.95$\cdot$10$^{4}$ &0.55 & 38 &$>$39 &  52  \\
3C\,445N  & $6.63   \cdot$10$^{5}$ &0.85 & 60 &  4.1 & 315  \\
3C\,445S  & $8.40   \cdot$10$^{5}$ &0.80 & 68 &  2.8 & 275  \\
&&&\\
\hline
\end{tabular}
\end{center}
\label{tab_synchro}
\end{table}

\begin{figure*}
\hspace*{-0.5cm}\rotatebox{0}{\includegraphics[width=6.36cm]{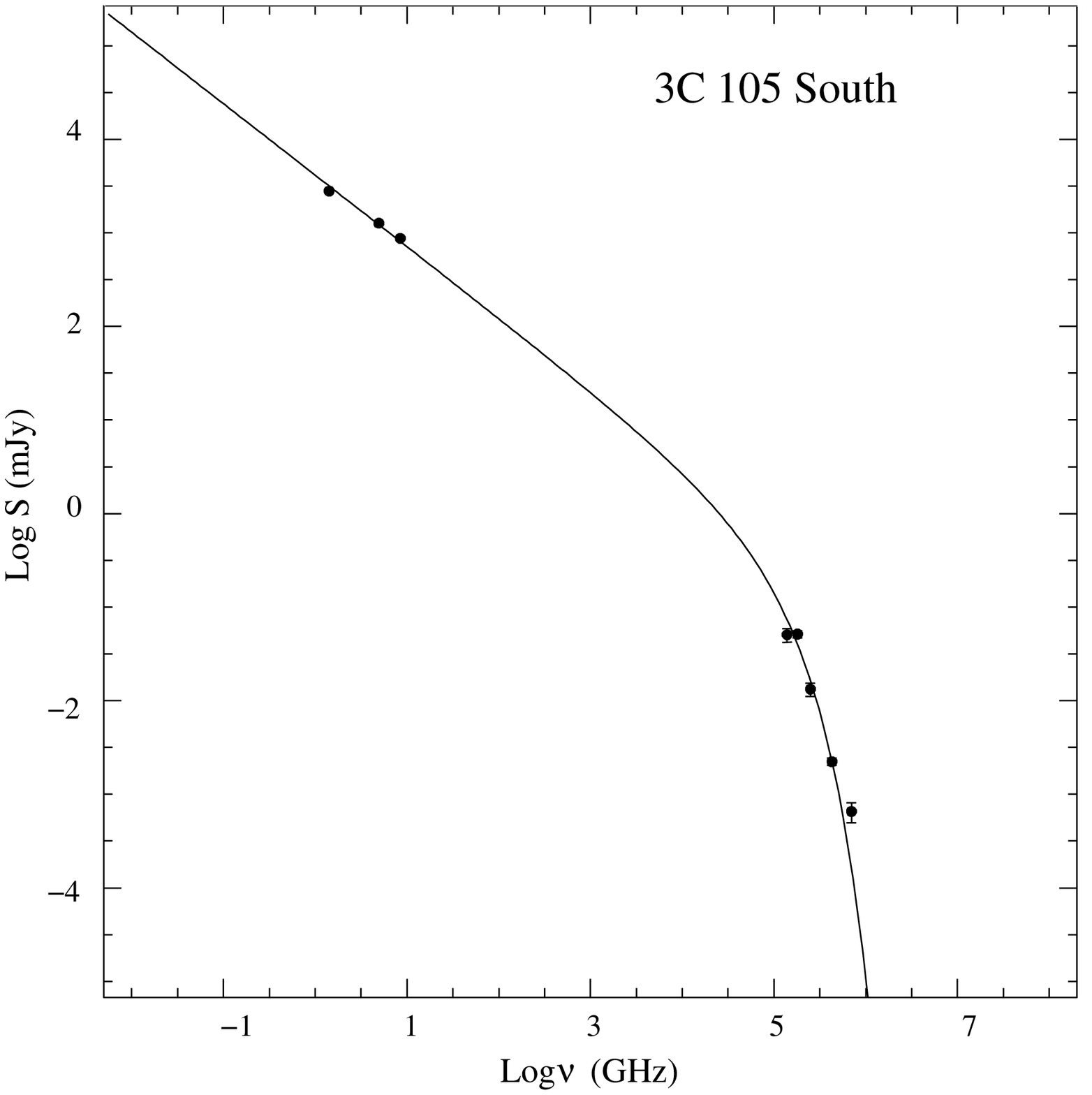}}
\hspace*{-0.5cm}\rotatebox{0}{\includegraphics[width=6.36cm]{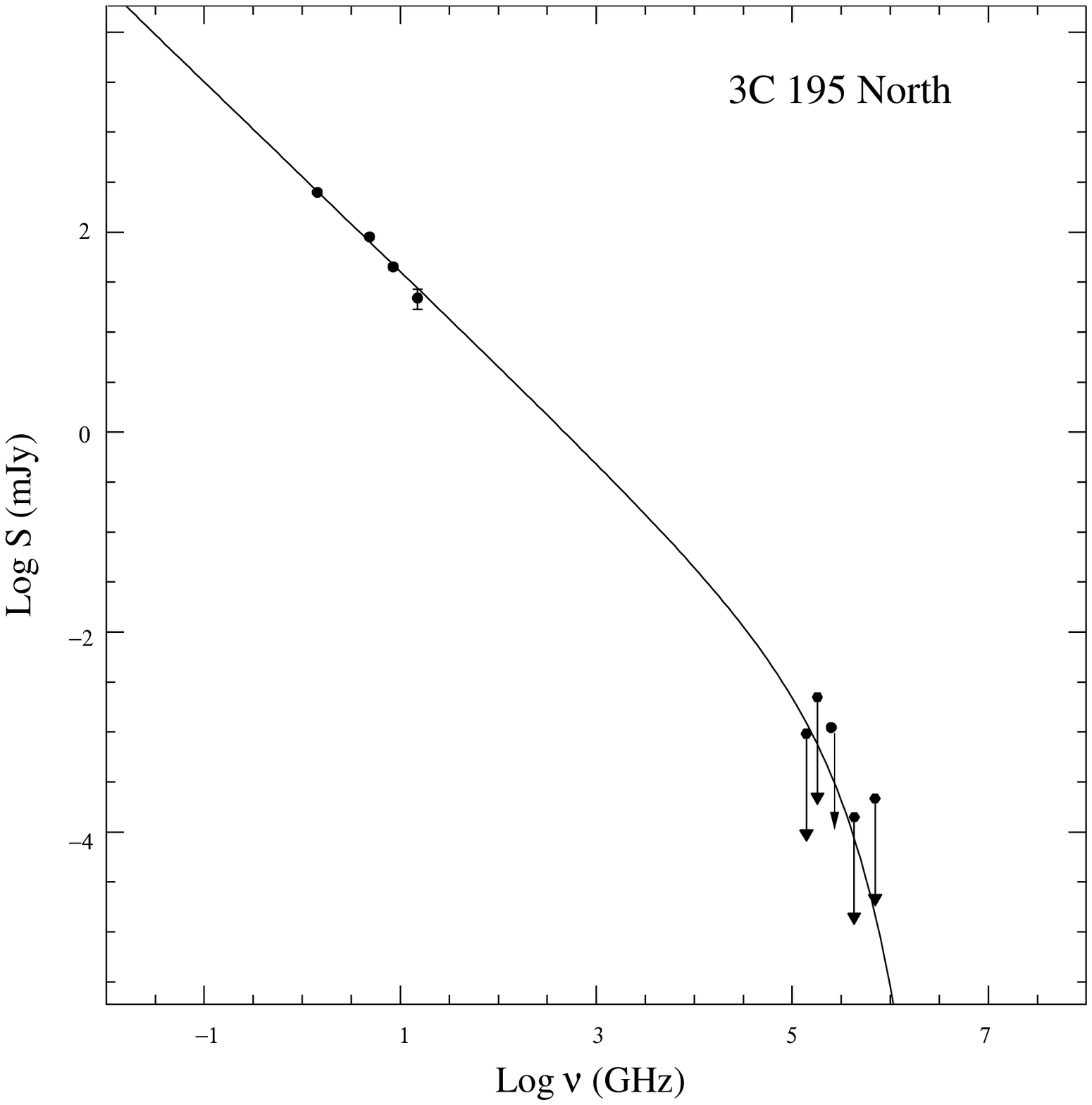}}
\hspace*{-0.5cm}\rotatebox{0}{\includegraphics[width=6.36cm]{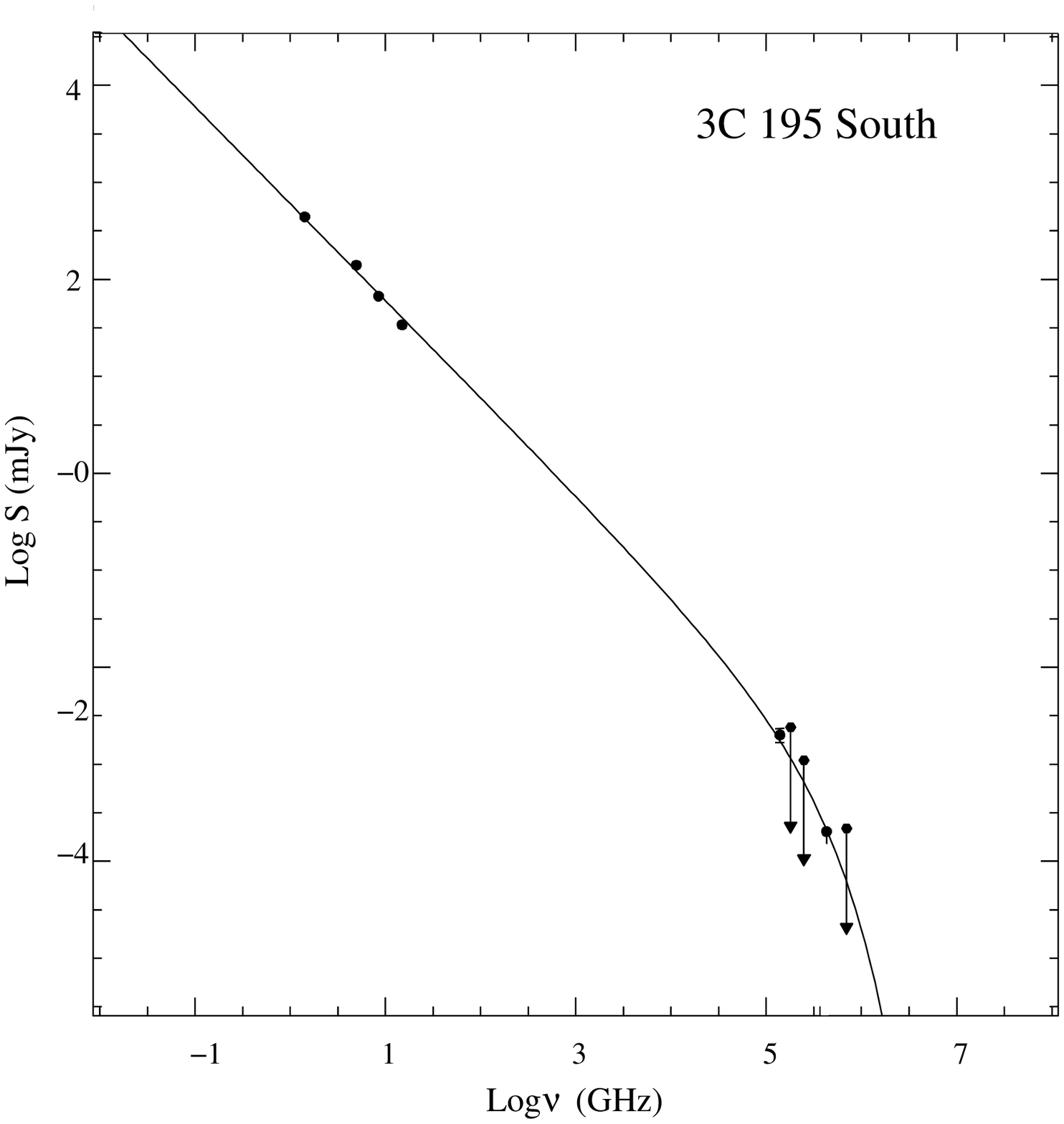}}
\hspace*{-0.5cm}\rotatebox{0}{\includegraphics[width=6.36cm]{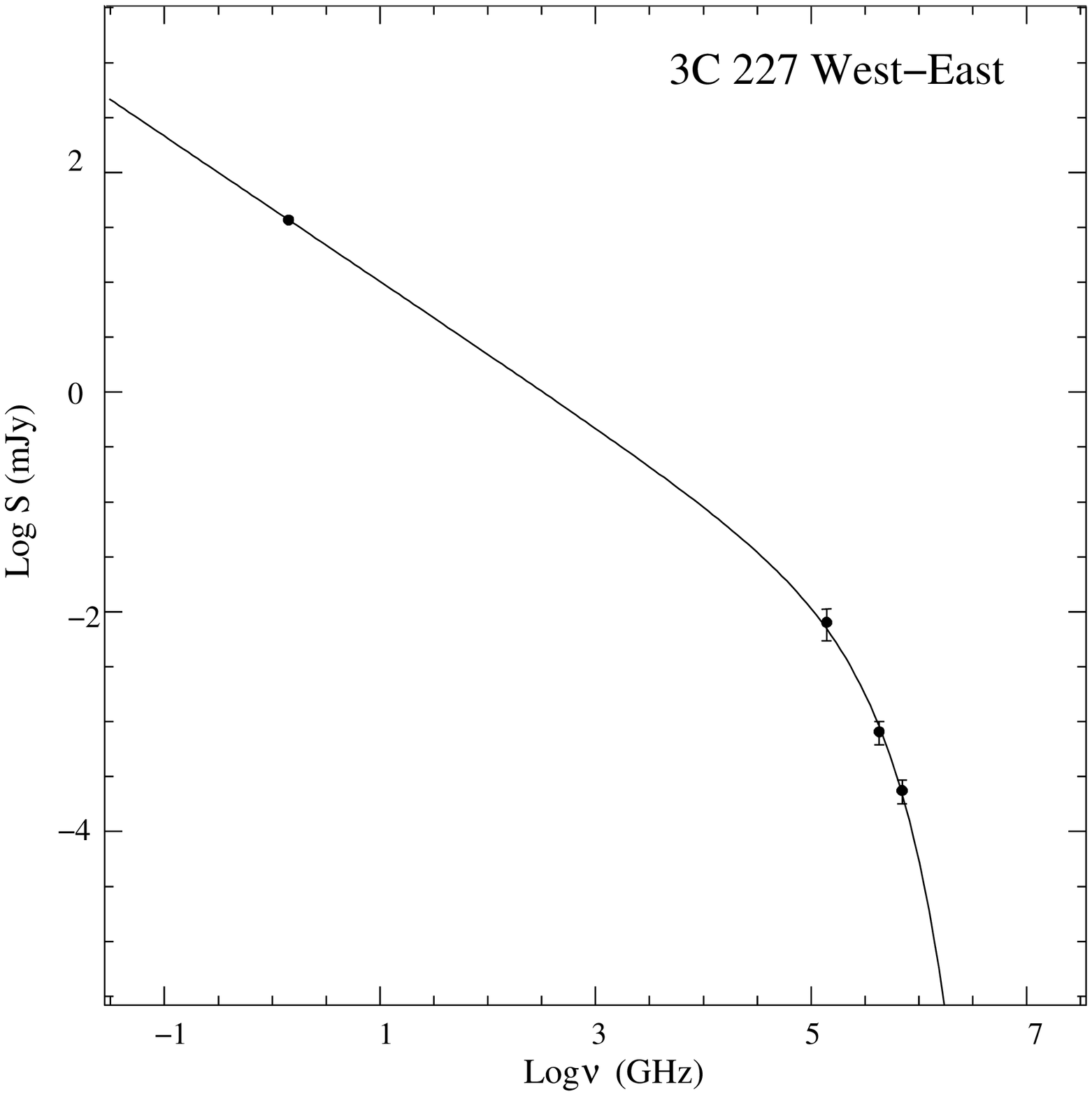}}
\hspace*{-0.5cm}\rotatebox{0}{\includegraphics[width=6.36cm]{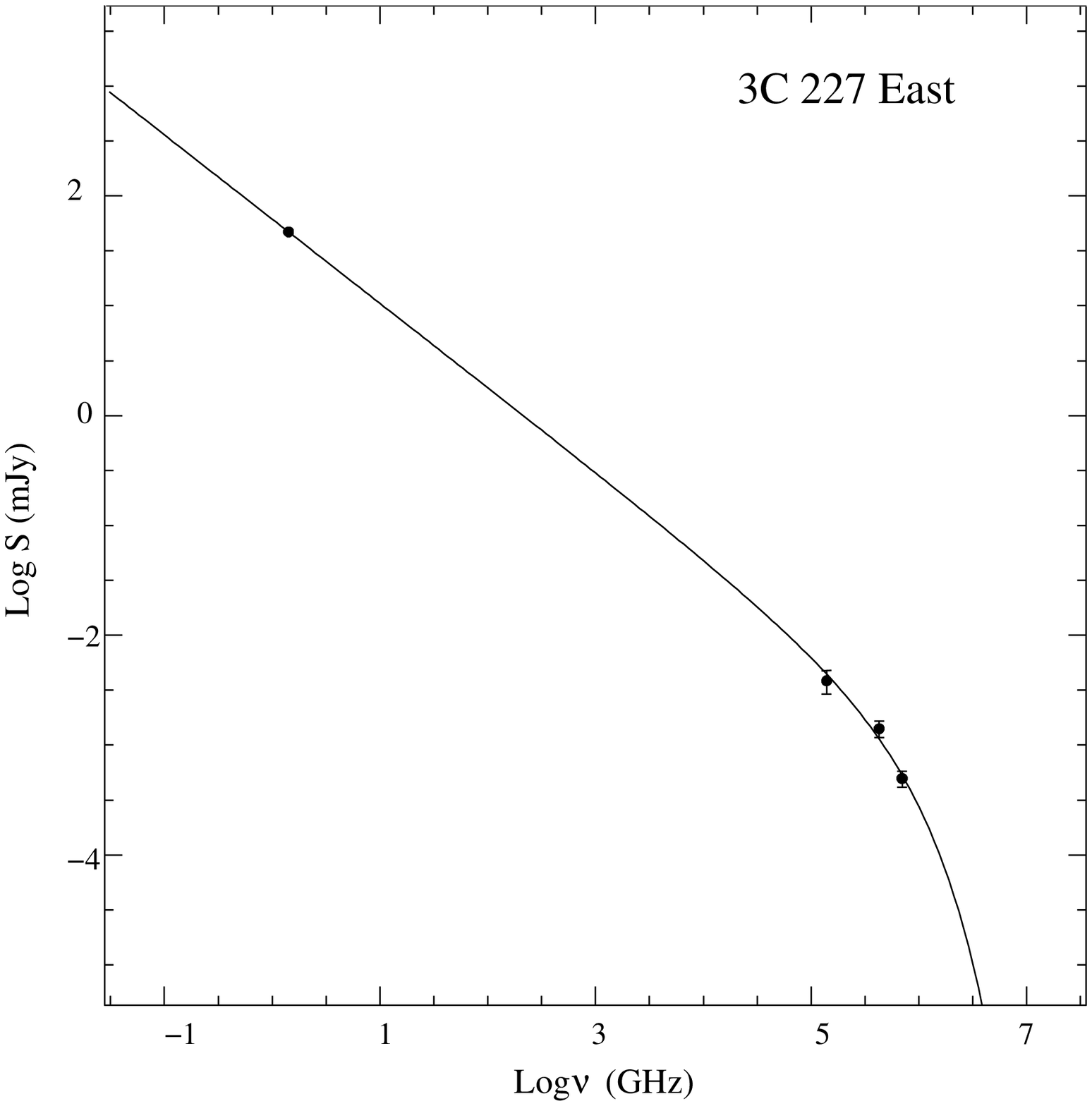}}
\hspace*{-0.5cm}\rotatebox{0}{\includegraphics[width=6.36cm]{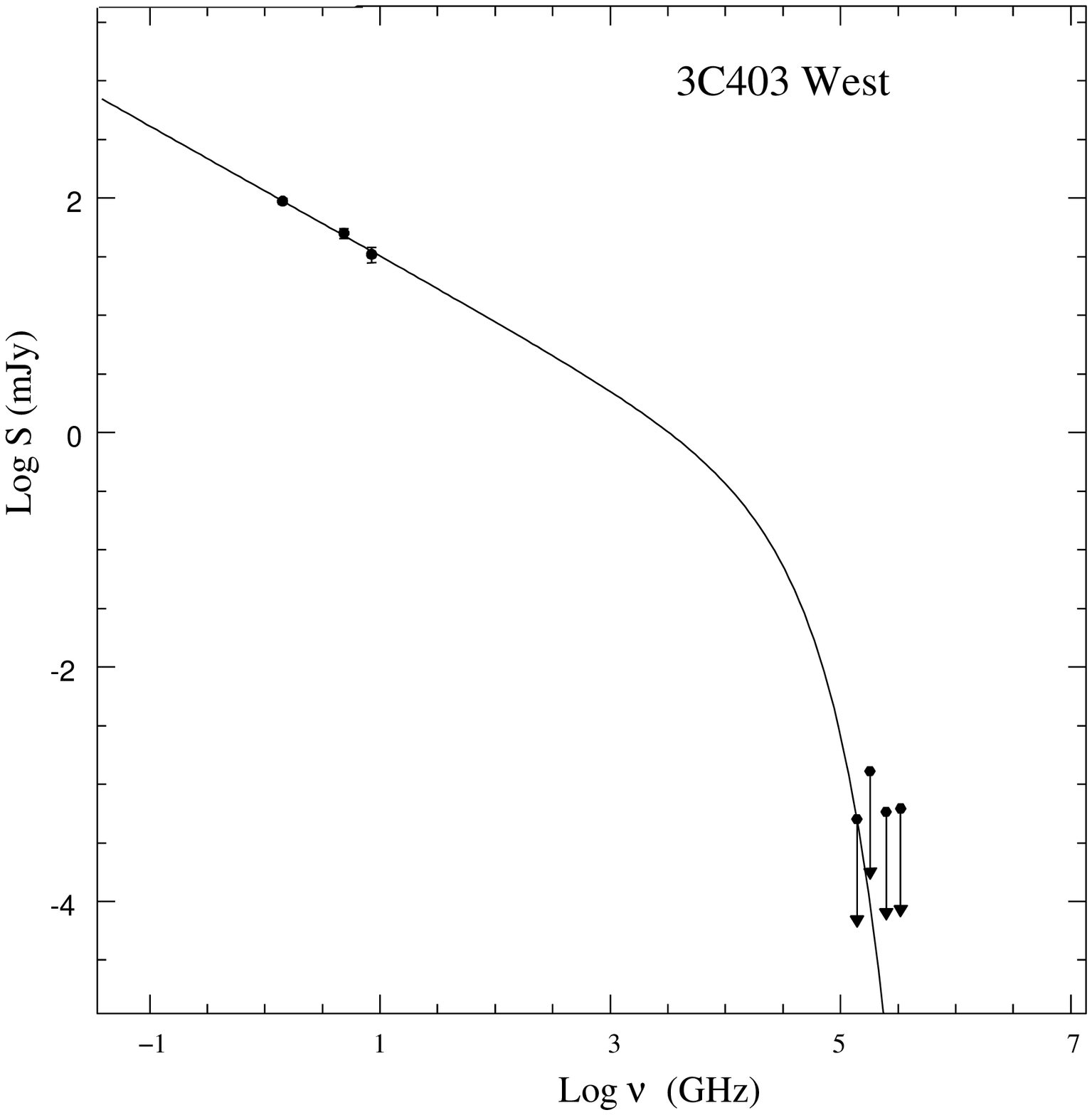}}
\rotatebox{0}{\includegraphics[width=6.36cm]{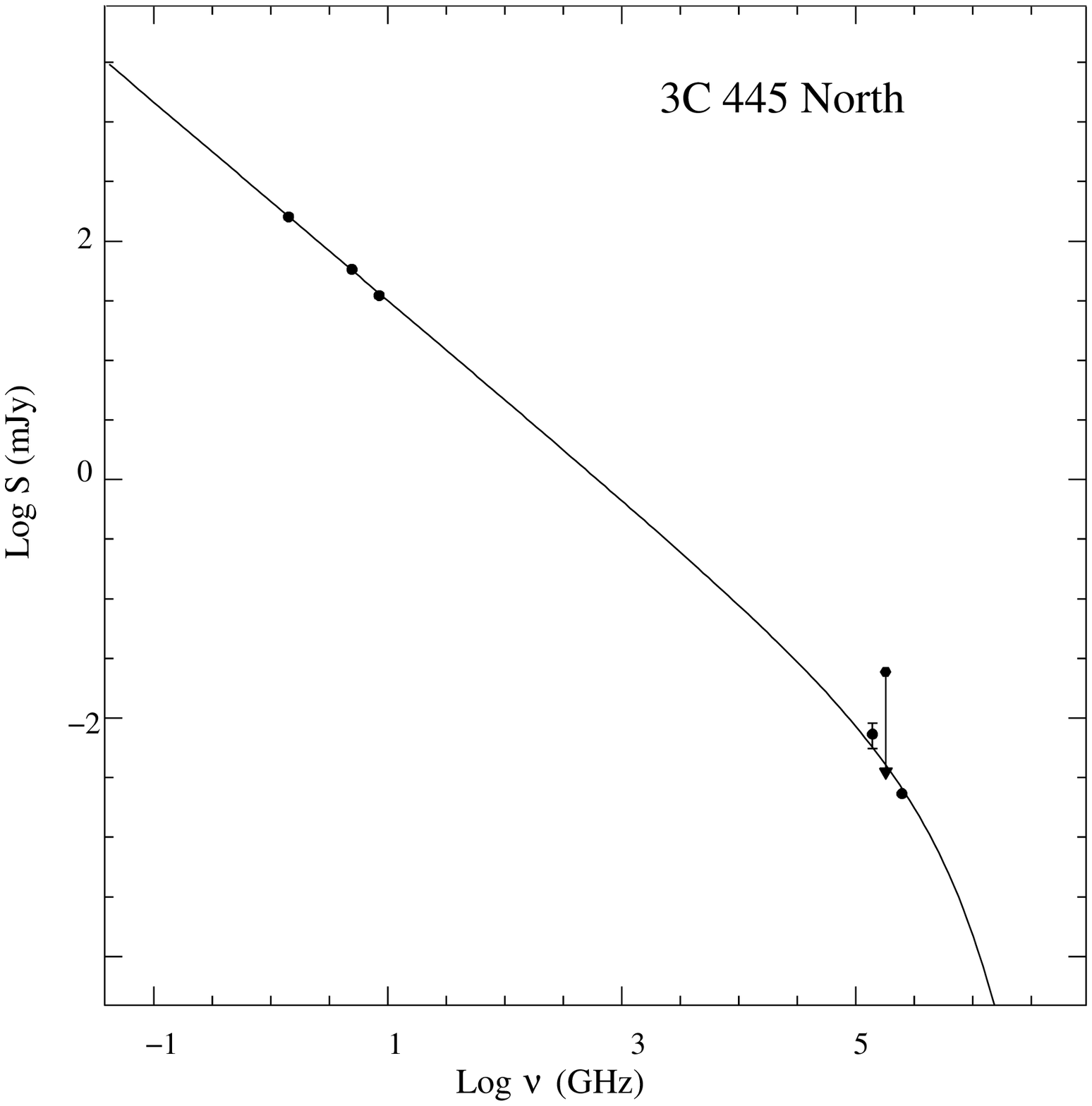}}
\rotatebox{0}{\includegraphics[width=6.36cm]{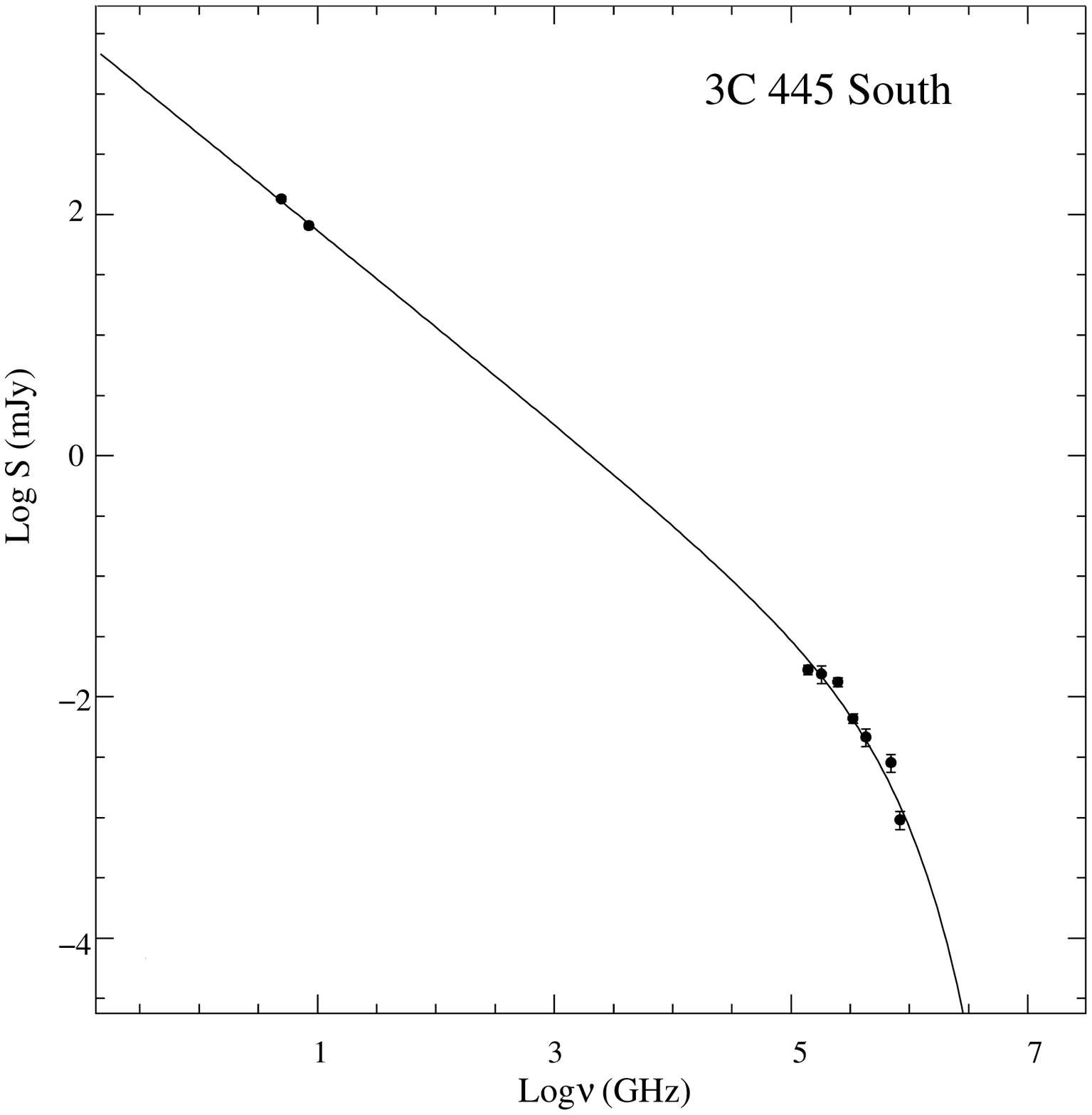}}
\caption{Synchrotron fits to the spectra of the 8 hotspots with sufficient
measurements. 3C\,327W was excluded as only one radio frequency and one upper
limit at NIR K-band is known.}
\end{figure*}

\section{Summary}

We present new data of a sample of low-power hotspots 
(P$_{\rm 1.4 GHz} \le 10^{25}$ W/Hz) observed at several NIR and
optical bands, complemented by high-resolution radio observations. From this
investigation we found:

\begin{itemize}

\item Of the 9 targeted hotspots at least 4 show a
clear detection at the NIR K-band, in several cases up to the optical (R, B)
bands. An additional two hotspots show NIR features which cannot yet safely 
be identified as the optical counterpart of the radio hotspot 
and will need deeper observations.
In any case this results in unprecedented detection rates between 45\% 
and 67\%. 

\item This high detection rate is attributed to the high break frequencies 
($10^5$ to $10^6$ GHz) of the continuum spectra of the detected sources
that emit the bulk of their synchrotron emission at optical wavelength.
High values of the break frequencies are indeed expected in the class of
low-power hotspots (Blundell et al. 1999; Brunetti et al. 2003).
For comparison, typical parameters of previously known optical hotspots
reported in the literature have equipartition fields $> 100\,\mu$G and 
break frequencies in the range of $10^{3}$ to $10^{4}$ GHz.

\item The large values of the synchrotron break frequencies in optically 
detected hotspots in our sample imply that the emitting
electrons have been injected in the hotspot volume less than $10^3$ years
ago.
A first obvious possibility to explain the lack of older electrons in the
emitting volume is an intermittent activity of the jet with a 
time-scale $\approx 10^3$ years.
A second possibility is that the emission from older electrons
in the hotspot volume is fading away due to fast adiabatic losses, in which
case
the optical emission is expected to be produced from compact (sub-kpc) 
and over-pressured knots.
Alternatively, in-situ Fermi-II mechanisms, with a particle re-acceleration
time scale $\approx 2-5\cdot 10^3$ years, partially balance the radiative 
losses of electrons. In this case, the radiative age from the measurement of 
the synchrotron break frequency should be considered a lower limit to the real 
age of the particles.

\item In almost all cases the detected NIR/optical emission is extended 
between 3 and 8 kpc. It breaks down in multiple substructure indicating that
we may be over-resolving the emission. The diffuse emission suggests that
in situ re-acceleration mechanisms may be at work in the hotspot volume,
as previously pointed out in 3C 445S (Meisenheimer et al. 1997,
Prieto et al. 2002). This would further suggest that the radiative ages
measured from the synchrotron break differ from the real ages of the emitting
particles.

\end{itemize}

\section*{Acknowledgements}

The National Radio Astronomy Observatory is a facility of the National
Science Foundation operated under cooperative agreement by Associated
Universities, Inc. This work has made use of the NASA/IPAC
Extragalactic Database NED which is operated by the JPL, Californian
Institute of Technology, under contract with the National Aeronautics
and Space Administration.

\end{document}